\documentclass[twocolumn,showpacs,preprintnumbers,superscriptaddress,aps,nofootinbib]{revtex4}

\usepackage[dvipdfmx]{graphicx}
\usepackage{dcolumn}
\usepackage{bm}
\usepackage{amsmath,amssymb,amsfonts}
\usepackage{latexsym}
\usepackage{bbm}
\usepackage{color}
\usepackage[colorlinks,
            linkcolor=black,
            anchorcolor=black,
            citecolor=black
            ]{hyperref}
\usepackage[utf8]{inputenc}

\begin{document}

\preprint{NCTS-PH/1724}
\preprint{ACFI-T17-16}

\title{
Standard Model with a Complex Scalar Singlet: Cosmological Implications and Theoretical Considerations}

\author{Cheng-Wei Chiang}%
\email{chengwei@phys.ntu.edu.tw}
\affiliation{Department of Physics, National Taiwan University, Taipei, Taiwan 10617, R.O.C.}
\affiliation{Institute of Physics, Academia Sinica, Taipei, Taiwan 11529, R.O.C.}
\affiliation{Physics Division, National Center for Theoretical Sciences, Hsinchu, Taiwan 30013, R.O.C.}
\affiliation{Kavli IPMU, University of Tokyo, Kashiwa, 277-8583, Japan}
\author{Michael J. Ramsey-Musolf}
\email{mjrm@physics.umass.edu}
\affiliation{Amherst Center for Fundamental Interactions, Department of Physics, University of Massachusetts-Amherst Amherst, MA 01003 USA}
\affiliation{California Institute of Technology, Pasadena, CA 91125 USA}
\author{Eibun Senaha}%
\email{senaha@ibs.re.kr}
\affiliation{Department of Physics, National Taiwan University, Taipei, Taiwan 10617, R.O.C.}
\affiliation{Center for Theoretical Physics of the Universe, Institute for Basic Science (IBS), Daejeon 34051, Korea}

\bigskip

\date{\today}

\begin{abstract}
We analyze the theoretical and phenomenological considerations for the electroweak phase transition and dark matter in an extension of the Standard Model with a complex scalar singlet (cxSM). In contrast with earlier studies, we use a renormalization group improved scalar potential and treat its thermal history in a gauge invariant manner. We find that the parameter space consistent with a strong first order electroweak phase transition (SFOEWPT) and present dark matter phenomenological constraints is significantly restricted compared to results of a conventional, gauge non-invariant analysis. In the simplest variant of the cxSM, recent LUX data and a SFOEWPT require a dark matter mass close to half the mass of the Standard Model-like Higgs boson. We also comment on various caveats regarding the perturbative treatment of the phase transition dynamics.
\end{abstract}


\maketitle

\section{Introduction}
\label{sec:intro}
Explaining the excess of matter over antimatter in the present Universe is a challenge and the interface of particle and nuclear physics with cosmology. It is well known that the Standard Model (SM) cannot account for the observed baryon asymmetry of the Universe (BAU)
\begin{eqnarray}
Y_B = \frac{n_B}{s} = (8.59 \pm 0.11) \times 10^{-11} \quad \textrm{(Planck)~\cite{Ade:2013zuv}}
\end{eqnarray}
where $n_B$ ($s$) is the baryon number (entropy) density. While the electroweak (EW) sphalerons of the SM fulfill the first of the ``Sakharov criteria''~\cite{Sakharov:1967dj},  the presence of baryon number (B) violating processes, the SM fails with regard to the remaining two requirements: sufficiently effective CP violation and out-of-equilibrium dynamics, assuming CPT symmetry is conserved. 

A plethora of baryogenesis scenarios beyond the Standard Model (BSM) have been proposed to remedy these SM shortcomings. Among the most theoretically attractive and phenomenologically testable is electroweak baryogenesis (EWBG)~\cite{Kuzmin:1985mm} (for reviews, see Refs.~\cite{Cohen:1993nk,Quiros:1994dr,Rubakov:1996vz,Funakubo:1996dw,Trodden:1998ym,Bernreuther:2002uj,Cline:2006ts,Morrissey:2012db,Konstandin:2013caa}). EWBG proceeds via bubble nucleation during a first-order electroweak phase transition (EWPT), providing the needed out-of-equilibrium conditions.  BSM CP-violating interactions at the bubble walls generate a net density of left-handed fermions, biasing the EW sphalerons into the creation of a non-zero baryon number density that diffuses into the expanding bubble interiors. For a sufficiently strong first-order electroweak phase transition (SFOEWPT), the sphaleron transitions in the bubble interiors are suppressed so as to preserve the generated baryon asymmetry.

The SM cannot accommodate a first-order EWPT since the observed Higgs boson is too heavy. Lattice computations indicate that the maximum mass for a first-order transition is 70-80 GeV~\cite{Kajantie:1996mn,Rummukainen:1998as,Csikor:1998eu,Aoki:1999fi}. An extended scalar sector from the SM, however, can allow a SFOEWPT for {a mass of} 125~GeV. The simplest such extensions involve the addition of gauge-singlet scalars. Such scalars may arise in a variety of contexts, such as the next-to-minimal supersymmetric Standard Model or U(1$)^\prime$ extensions of the SM. By focusing on the role played by the associated singlet scalars, one may infer general features of the EWPT that are common to these and other scenarios without making reference to other model-specific details. 

The possibilities for a SFOEWPT with an additional real singlet scalar (dubbed the ``xSM'' \cite{Barger:2007im}), along with its phenomenological consequences for collider studies, have been studied extensively~\cite{Profumo:2007wc,Ahriche:2007jp,Noble:2007kk,Cline:2009sn,Espinosa:2011ax,No:2013wsa,Alanne:2014bra,Profumo:2014opa,Kotwal:2016tex,Vaskonen:2016yiu,Huang:2017jws,Chen:2017qcz}. The xSM implies the existence of two neutral mass eigenstates, $H_{1,2}$, that are mixtures of the neutral doublet and real singlet. Signatures associated with a SFOEWPT include reduced SM-like Higgs signal strengths, modifications of the Higgs trilinear self-coupling, exotic decays of the SM-like Higgs boson, and resonant di-Higgs production. The last provides a particularly interesting opportunity not only for the LHC~\cite{No:2013wsa,Huang:2017jws} but also for a possible future higher-energy $pp$ collider~\cite{Kotwal:2016tex,Chen:2017qcz}. Under some conditions, when the xSM assumes a $\mathbb{Z}_2$ symmetry, it may also provide a dark matter (DM) candidate with an associated vector boson fusion signature for the next-generation $pp$ collider\cite{Curtin:2014jma}.

In this work, we focus on the extension with one complex scalar singlet, the cxSM. This scenario was introduced in Ref.~\cite{Barger:2008jx}, where it was shown that one could obtain both the conditions for a SFOEWPT and a viable DM candidate and  even in the absence of a $\mathbb{Z}_2$ symmetry. The presence of a global U(1) symmetry that is both spontaneously and softly broken implies the existence of two mixed states {$H_{1,2}$} of the xSM plus a pseudoscalar dark matter candidate $A$. A subsequent extensive study of the implications for vacuum stability and DM phenomenology was given in Ref.~\cite{Gonderinger:2012rd}. A less minimal version of the cxSM was also studied recently by the authors of Ref.~\cite{Jiang:2015cwa} who also included a non-renormalizable, CP-violating top quark-scalar interactions as needed to generate the BAU. For suitable choices of the model parameters, it appears possible to obtain both the observed BAU and DM relic density while satisfying present phenomenological constraints.~\footnote{Ref.~\cite{Sannino:2015wka} also considers a complex singlet extension that includes an additional neutral fermion as the dark matter candidate.}

Given the richness of this simple scenario, a more thorough study of open theoretical issues as well as phenomenological implications is in order.  These issues include: 
\begin{itemize}
\item To what extent do the conclusions of earlier studies hold when an appropriately gauge-invariant treatment of EWPT properties is performed?
\item To what extent can the renormalization scale-dependence of computed EWPT properties be controlled?
\item What can one conclude about the possible thermal history of EWSB in this scenario when a gauge-invariant and scale-invariant treatment is performed?
\item To what extent does the gauge-invariant scale ${\bar v}(T)$ \cite{Patel:2011th} associated with electroweak symmetry breaking (EWSB) characterize the sphaleron energy $E_\mathrm{sph}(T)$? Is the latter simply proportional to the former? 
\end{itemize}

In addressing these issues, we perform a gauge-invariant (GI) study of the cxSM EWPT dynamics following the framework of Ref.~\cite{Patel:2011th} and show how use of a renormalization group (RG)-improved version of the model significantly reduces the dependence on the renormalization scale. We also delineate various possibilities of the cxSM thermal history and analyze the effectiveness of baryon number preservation as a function of model parameters. Lastly, we connect these features to phenomenological signatures. We find that the null results for spin-independent DM-nucleus scattering obtained by the LUX collaboration~\cite{Akerib:2016vxi} severely constrain the SFOEWPT-viable parameter space. We can say with confidence that only the finely-tuned region for $m_A\approx m_h/2$ remains viable, where $m_A$ and $m_h$ are the pseudoscalar dark matter and Higgs boson masses, respectively. While the parameter space for much larger values of $m_A$ (approaching 1 TeV) may also accommodate the LUX bounds while providing for a SFOEWPT, the remaining theoretical ambiguities associated with perturbative treatments of the EWPT dynamics render this possibility less certain. Our conclusions in this respect are less optimistic than those given in Ref.~\cite{Jiang:2015cwa} that carried out a gauge-dependent treatment of the scalar effective potential. Although the latter study included a wider set of terms in the scalar potential that we consider below, a comparison of our GI, RG-improved analysis with the conventional treatment suggests that the SFOEWPT-viable parameter space associated with the wider set of operators used in Ref.~\cite{Jiang:2015cwa} may be more restricted { than} indicated in that study.

The discussion of our analysis is organized as follows. In Section \ref{sec:model}, we introduce the scalar potential of the cxSM and analyze EWSB at $T=0$. In Section \ref{sec:thermal} we delineate the various possibilities for the thermal history of EWSB in the model and  introduce two representative scenarios. Section \ref{sec:EWBG} gives a detailed discussion of the conditions needed for successful EWBG, a.k.a. a SFOEWPT: sufficiently rapid bubble nucleation and sufficiently efficient baryon number preservation inside the broken phase bubbles. In Section \ref{sec:girge} we review the methods for performing a GI analysis and RG improvement. Section \ref{sec:numerics} contains an extensive numerical study of the two representative EWSB scenarios given in Section \ref{sec:thermal}. In Sections \ref{sec:DM} and \ref{sec:pheno} we then apply the constraints from DM phenomenology. We conclude in Section \ref{sec:conclude}.


\section{Model}
\label{sec:model}
Consider the SM extended with a complex gauge-singlet scalar field (cxSM).  
Following Ref.~\cite{Barger:2008jx}, we analyze a simplified version of the Higgs potential:
\begin{align}
\lefteqn{V_0(H,\mathbb{S})} \nonumber\\
&= \frac{m^2}{2}H^\dagger H+\frac{\lambda}{4}(H^\dagger H)^2
	+\frac{\delta_2}{2}H^\dagger H|\mathbb{S}|^2+\frac{b_2}{2}|\mathbb{S}|^2+\frac{d_2}{4}|\mathbb{S}|^4 \nonumber\\
&\quad
	+\bigg[a_1\mathbb{S}+\frac{b_1}{4}\mathbb{S}^2+{\rm h.c.}\bigg] ~.
\label{eq:vhs}
\end{align}
where $\mathbb{S}$ is the complex singlet scalar.  The terms in the first line on the right-hand side (RHS) of Eq.~(\ref{eq:vhs}) are invariant under the global U(1) transformation $\mathbb{S}\to\exp{(i\theta)} \mathbb{S}$. The remaining terms on the RHS explicitly break the global U(1) symmetry. The coefficients 
$a_1$ and $b_1$ are generally complex. While {one} of them can be made real through the field redefinition of $\mathbb{S}$, in general $\mathrm{Im} (b_1^\ast a_1^2)$ is non-vanishing. The existence of a viable DM candidate requires that one set the re-phasing invariant $\mathrm{Arg} (b_1^\ast a_1^2)=0$. Note that one may include additional renormalizable operators that break the global U(1)~\cite{Barger:2008jx}.  But for simplicity, we have only included a minimal subset that closes under renormalization. The study of Ref.~\cite{Jiang:2015cwa} includes a subset of these additional operators. 

In the limit of vanishing $a_1$ and $b_1$, the occurrence of a non-zero vacuum expectation value (VEV) of $\mathbb{S}$, $\langle \mathbb{S}\rangle = v_S$, will break the global U(1), giving rise to a massless Nambu-Goldstone mode, $A$. The $a_1$ and $b_1$ terms induce a non-zero $A$ mass without sacrificing its stability, thereby making it a viable cold DM candidate. In the limit of vanishing $a_1$, $V(H,\mathbb{S})$ also possesses a $\mathbb{Z}_2$ symmetry that would be spontaneously broken by a non-vanishing $v_S$. The explicit $\mathbb{Z}_2$-breaking $a_1$ term thus eliminates any potentially problematic domain walls (see, {\em e.g}, \cite{Abel:1995wk} and references therein). After electroweak symmetry breaking (EWSB), where $\langle H^0\rangle=v$, the U(1)-symmetric $\delta_2$ term will lead to mixing between the SM Higgs boson and the remaining degree of freedom in $\mathbb{S}$, leading to the two mass eigenstates $H_{1,2}$. 

For concreteness, we choose the following representations for the scalar fields after spontaneous symmetry breaking (SSB): 
\begin{align}
H(x) &=
	\left(
		\begin{array}{c}
		G^+(x) \\
		\frac{1}{\sqrt{2}}\big(v_0+h(x)+iG^0(x)\big)
		\end{array}
	\right) ~,\\
\mathbb{S}(x) &=\frac{1}{\sqrt{2}}\big(v_{S0}+S(x)+iA(x)\big) ~,
\end{align}
where $v_0(\simeq246~{\rm GeV})$ and $v_{S0}$ are the VEV's at zero temperature, and $G^{0,\pm}$ are the Nambu-Goldstone bosons. 
For a recent analysis of the Higgs phenomenology of this model and a comparison with the Next-to-Minimal Supersymmetric Standard Model, see Ref.~\cite{Costa:2015llh} (for a vacuum stability analysis at  the two-loop level, see also Ref.~\cite{Costa:2014qga}.)
In the current investigation, all the complex phases that can mix $S(x)$ with $A(x)$ 
are assumed to be zero, and $A(x)$ is  assumed to be the DM candidate.

The tadpole conditions of $V_0$ for $h$ and $S$ are respectively 
\begin{align}
\left\langle\frac{\partial V_0}{\partial h}\right\rangle 
&= v_0\left[\frac{m^2}{2}+\frac{\lambda}{4}v_0^2+\frac{\delta_2}{4}v^2_{S0}\right]=0 ~, 
\label{tree-tad_h}\\
\left\langle\frac{\partial V_0}{\partial S}\right\rangle
&= v_{S0}\left[\frac{b_2}{2}+\frac{\delta_2}{4}v_0^2
	+\frac{d_2}{4}v^2_{S0}+\sqrt{2}\frac{a_1}{v_{S0}}+\frac{b_1}{2}\right] = 0 ~,
	\label{tree-tad_hS}
\end{align}
where the symbol $\langle\cdots\rangle$ denotes that the fluctuating fields are taken to be zero.

For the $v_{S0}\neq0$ case, after imposing Eqs.~(\ref{tree-tad_h}) and (\ref{tree-tad_hS}), the mass matrix of the CP-even bosons takes the form
\begin{align}
\mathcal{M}_S^2 &= 
\left(
\begin{array}{cc}
	\frac{\lambda}{2}v_0^2 & \frac{\delta_2}{2}v_0v_{S0} \\
	\frac{\delta_2}{2}v_0v_{S0} & \frac{d_2}{2}v_{S0}^2-\frac{\sqrt{2}a_1}{v_{S0}}
\end{array}
\right) ~,
\label{Ms_tree}
\end{align}
which can be diagonalized by an orthogonal matrix $O$:
\begin{align}
O^T\mathcal{M}_S^2O  =
\left(
	\begin{array}{cc}
	m_{H_1}^2 & 0 \\
	0 & m_{H_2}^2
	\end{array}
\right) ~, \quad
O(\alpha) = 
\left(
	\begin{array}{cc}
	c_\alpha & -s_\alpha \\
	s_\alpha & c_\alpha
	\end{array}
\right) ~,
\label{diagonalization}
\end{align}
where $\alpha$ is the mixing angle such that $(h,S)^T= O(\alpha)(H_1, H_2)^T$.
In our study, we consider a case in which $H_1(H_2)$ is the doublet (singlet)-like Higgs boson
by focusing on the range, $-\pi/4\le \alpha \le \pi/4$.

In our study, we fix $m_{H_1}=125$ GeV.
Under this assumption, $\lambda$ ($d_2$) for a nonzero $\alpha$ is always greater (smaller) than that in the $\alpha=0$ case. Similarly, the CP-odd scalar mass is given by
\begin{align}
m_A^2= -\frac{\sqrt{2}a_1}{v_{S0}}-b_1.\label{mA_tree}
\end{align}
Using Eqs.~(\ref{tree-tad_h})-(\ref{mA_tree}), one can trade off some of the original parameters with physically relevant quantities. Explicitly, we take the following as the input parameters: $v_0$, $v_{S0}$, $m_{H_1}(=125$ GeV), $m_{H_2}$, $\alpha$, $m_A$ and $a_1$.

For the $v_{S0}=0$ and $a_1=0$, on the other hand, one has
\begin{align}
\begin{split}
m_{H_1} ^2&= \frac{\lambda }{2}v_0^2, \quad m_{H_2}^2 = \frac{b_1+b_2}{2} + \frac{\delta_2}{4}v_0^2,\\
m_A^2&=\frac{-b_1+b_2}{2} + \frac{\delta_2}{4}v_0^2. \label{Hmass_S2}
\end{split}
\end{align}
The input parameters in this case are $v_0$, $m_{H_1}(=125$ GeV), $m_{H_2}$, $m_A$, $\delta_2$ and $d_2$. 


\section{Thermal history} 
\label{sec:thermal}
The behavior of the potential $V(H,\mathbb{S})$ at non-zero temperature, $T$, can lead to different patterns of symmetry breaking, depending on the choice of model parameters. To gain some intuition for various possibilities, we consider the impact of the thermal mass contributions that are responsible for symmetry restoration at high-$T$. For simplicity, we will begin by setting $a_1=0$. We will return to a discussion of the full set of thermal loop contributions later. The ``high-temperature effective potential'' in this case is given by
\begin{align}
V^{{\rm high\mathchar`-}T}(\varphi, \varphi_S; T)
&=V_0(\varphi, \varphi_S)+\frac{1}{2}(\Sigma_{H}\varphi^2+\frac{1}{2}\Sigma_{S}\varphi_S^2) T^2\ ,
\label{V_highT_cxSM}
\end{align}
where $\varphi$ and $\varphi_S$ denote the neutral doublet and singlet background fields, respectively, and 
\begin{align}
\Sigma_{H}
= \frac{\lambda}{8}+\frac{\delta_2}{24}+\frac{3g_2^2+g_1^2}{16}+\frac{y_t^2}{4},\quad 
\Sigma_{S}
=\frac{\delta_2+d_2}{12} ~,
\end{align}
where $g_1$ and $g_2$ denote, respectively, the U(1)$_Y$ and SU(2)$_L$ gauge couplings.
%
\begin{figure}[t]
\center
\includegraphics[width=3.5cm]{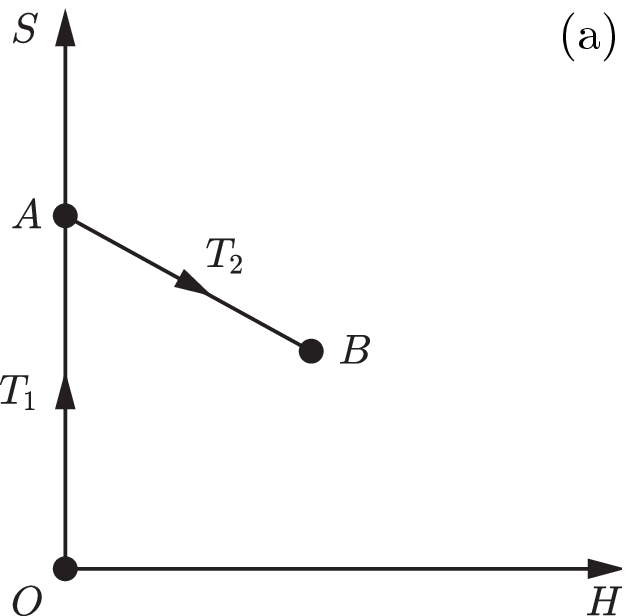}
\includegraphics[width=3.5cm]{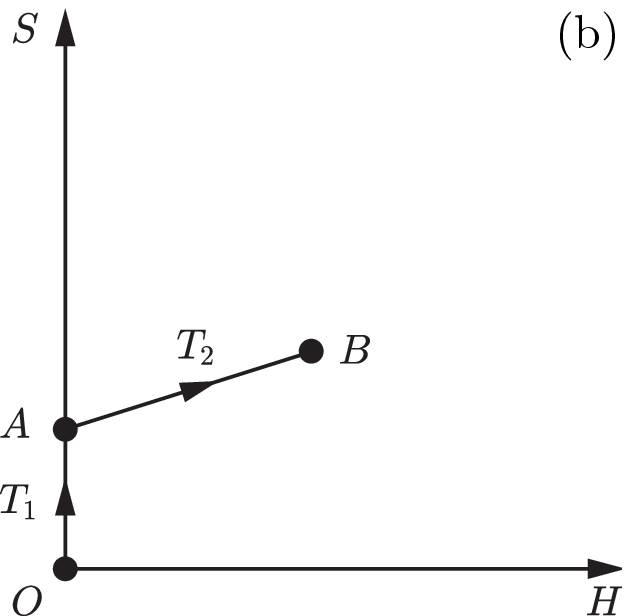}
\includegraphics[width=3.5cm]{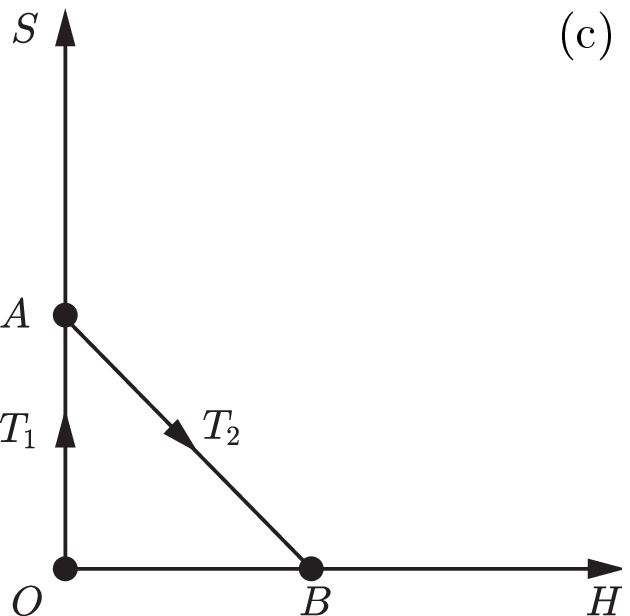}
\includegraphics[width=3.5cm]{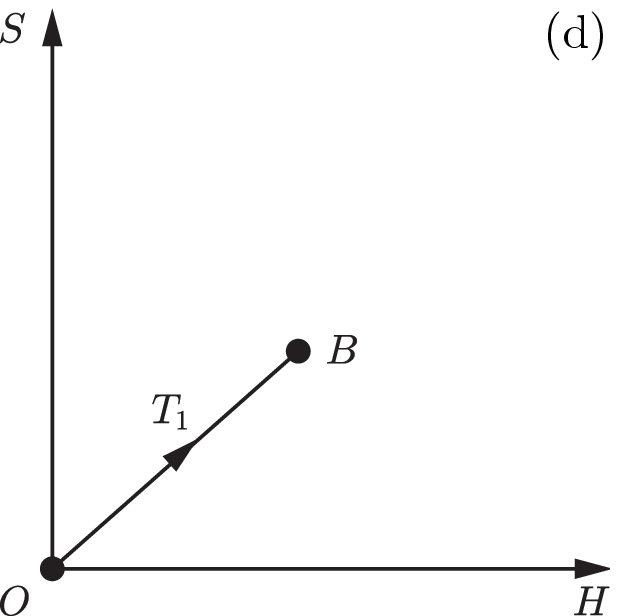}
\caption{Patterns of symmetry breaking at finite temperature for $a_1=0$. For cases (a) and (b), one has $\delta_2>0$ and $\delta_2<0$, respectively. For case (c), the singlet VEV is nonzero at $T\neq0$ while zero at $T=0$. For case (d), the EWPT occurs in one step.
}
\label{fig:PTs}
\end{figure}
%

The various possibilities for the thermal history of EWSB are illustrated in Fig.~\ref{fig:PTs}. The transition to the present \lq\lq Higgs phase" vacuum may occur either in two-steps [panels (a-c)] or a single step [panel (d)]. EWSB at $T=0$ requires $m^2<0$ but $b_2$ may have either sign. For $b_2>0$, the only relevant impact of the singlet scalars is via thermal loops. As shown in Refs.~\cite{Espinosa:2007qk,Noble:2007kk}, one requires a large number of additional singlet scalars to yield a SFOEWPT in this case. Consequently, we will focus on scenarios where $m^2<0$ and $b_2<0$, for which $T=0$ minima will exist for both $v_0$ and $v_{S0}$ non-zero. 

In the limit of vanishing $a_1$, the only minimum of the theory at sufficiently high-$T$ occurs at the origin,
denoted by ``O''. As $T$ decreases, one generically expects that a secondary minimum at $\varphi_S\equiv \bar{v}_S^{\rm A}\not=0$ will first appear, since $\Sigma_S< \Sigma_H$. At a temperature $T_1$, the minimum at $\bar{v}_S^{\rm A}$ will become the global minimum, indicated by ``A'' in Fig.~\ref{fig:PTs}. As $T$ further decreases, an additional  minimum at ($\varphi\equiv \bar{v}\not=0$, $\varphi_S\equiv \bar{v}_S^{\rm B}\not=0$) develops, becoming the global minimum at temperature $T_2 < T_1$, corresponding to point \lq\lq B" in Fig.~\ref{fig:PTs}. The Universe will then follow a two-step symmetry-breaking trajectory in the field space shown in Fig.~\ref{fig:PTs}, where one may have either $\bar{v}_S^{\rm A}(T_2) - \bar{v}_S^{\rm B}(T_2) > 0$ or $< 0$. We will henceforth denote $T_2$ as the EWSB critical temperature, $T_C$, and the value of $\varphi$ at this temperature as $\bar{v}(T_C)$. After a straightforward calculation, one finds
\begin{align}
{\bar v}(T_C)&\simeq \sqrt{\frac{2\delta_2\bar{v}_S^{\rm A}(T_C)}{\lambda }
	\big(\bar{v}_{S}^{\rm A}(T_C)-\bar{v}_{S}^{\rm B}(T_C)\big)} ~,\label{vc} \\
T_C&\simeq \sqrt{\frac{1}{2\Sigma_H}\left(-m^2-\frac{(\bar{v}_S^{\rm A}(T_C))^2}{2}\delta_2\right)} ~.
\label{Tc}
\end{align}
Here, the bar over $v(T_C)$ indicates that it has been computed using the high-$T$ effective potential given in Eq.~(\ref{V_highT_cxSM}).
$T_C$ and $\bar{v}(T_C)$ obtained in this way are the leading-order gauge-invariant results. 
For positive (negative) $\delta_2$ one has $\bar{v}_{S}^{\rm A}(T_C)$ larger (smaller) 
than $\bar{v}_{S}^{\rm B}(T_C)$.  
In addition, for positive $\delta_2$, the potential will have a potential barrier between the minima at A and B. In this case, the EWPT at $T=T_C$ is first-order. 
Note, however, that a sufficiently large, positive $\delta_2$ may render B at $T=0$ metastable, since the energy difference between phase A and phase B  can become negative.
Therefore, there should be an upper bound on the magnitude of $\delta_2$, as will be discussed below.
For negative $\delta_2$, in contrast, $T_C$ is always raised to prevent
$\bar{v}(T_C)/T_C$ from becoming sizeable. 
In fact,  our numerical analysis (see below) does not yield a SFOEWPT for $\delta_2<0$. 

Alternately, the EWSB may occur directly from the origin to point B, as shown by Type-(d) in Fig.~\ref{fig:PTs}. In the high-$T$ effective theory, this transition is not first order, since $\bar{v}_S^{\rm A}$ is zero, as seen from Eq.~(\ref{vc}). However, the additional thermal corrections appearing in $V_1(\boldsymbol{\varphi}; T)$ (defined below) can generate a thermal barrier that, in principle, may accommodate a first order transition. 

Using the high-$T$ potential with the thermal cubic term of $S$, one  finds~\cite{Barger:2008jx}
\begin{align}
\frac{v(T_C)}{T_C} 
\simeq \frac{8E_S}{\lambda+2 {\delta_2} \tan^2\gamma(T_C)+d_2\tan^4\gamma(T_C)},
\label{1stepPT}
\end{align}
where $\tan\gamma(T_C)=v_S(T_C)/v(T_C)$ 
and $E_S$ denotes the coefficient of the thermal cubic term of $S$.
In this case, as opposed to the aforementioned two-step EWPT cases, 
SFOEWPT is strengthened by the {\it negative} $\delta_2$ term. 
However, it should be emphasized that $E_S$ would be highly suppressed if the singlet Higgs mass
is dominated by the mass parameters $b_1$ and $b_2$, as in the Higgs thermal loop.

In our numerical study below, we do not find any SFOEWPT-viable parameter choices for this one-step transition. In addition, inclusion of additional U(1)-breaking cubic operators, such as $H^\dagger H\mathbb{S}$ and/or $\mathbb{S}^3$ may also allow-for a SFOEWPT~\cite{Profumo:2007wc,Jiang:2015cwa} as they contribute to the numerator in Eq.~(\ref{1stepPT}).

\begin{figure}[t]
\center
\includegraphics[width=3.5cm]{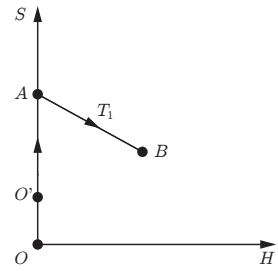}
\caption{Symmetry breaking at finite temperature for $a_1\neq 0$.
In this case the $\text{O}\to\text{A}$ transition is absent, and the initially nonzero $\bar{v}_S^{\rm A}(T)$ 
smoothly increases until the temperature reaches $T_1$ where EWPT happens.
}
\label{fig:PT_nonzeroa1}
\end{figure}
%

The situation can be more complex when taking  $a_1\not=0$ and including the remaining zero-temperature and thermal loop effects  encoded in the one-loop effective potential:
\begin{equation}
V_\mathrm{eff}(\boldsymbol{\varphi}; T) = V_0(\boldsymbol{\varphi}) + V_1(\boldsymbol{\varphi}; T) 
\label{Veff}
\end{equation}
where  $\boldsymbol{\varphi} = (\varphi, \varphi_S)$,
\begin{align}
V_1(\boldsymbol{\varphi}; T) = \sum_{j}n_j
\bigg[
V_{\rm CW}(\bar{m}_j^2)
	+\frac{T^4}{2\pi^2}I_{B,F}\left(\frac{\bar{m}_j^2}{T^2}\right)
\bigg] ~,\label{V1}
\end{align}
$n_j$ counts the degrees of freedom for particle species $j$, and $\bar{m}_j$ are $\boldsymbol{\varphi}$-dependent masses.  The Coleman-Weinberg potential $V_{\rm CW}$ and $I_{B,F}(a^2)$ are respectively given by~\cite{Coleman:1973jx,Dolan:1973qd}
\begin{align}
V_{\rm CW}(m^2) 
&= \frac{m^4}{64\pi^2}\left(\ln\frac{m^2}{\mu^2}-c\right) ~, \\
I_{B,F}(a^2) 
&= \int_0^\infty dx~x^2\ln\Big(1\mp e^{-\sqrt{x^2+a^2}}\Big) ~,
\end{align}
where $c=3/2$ for scalars and fermions and $5/6$ for gauge bosons, and $\mu$ is the renormalization scale. 

\begin{figure*}[t]
\center
\includegraphics[width=7cm]{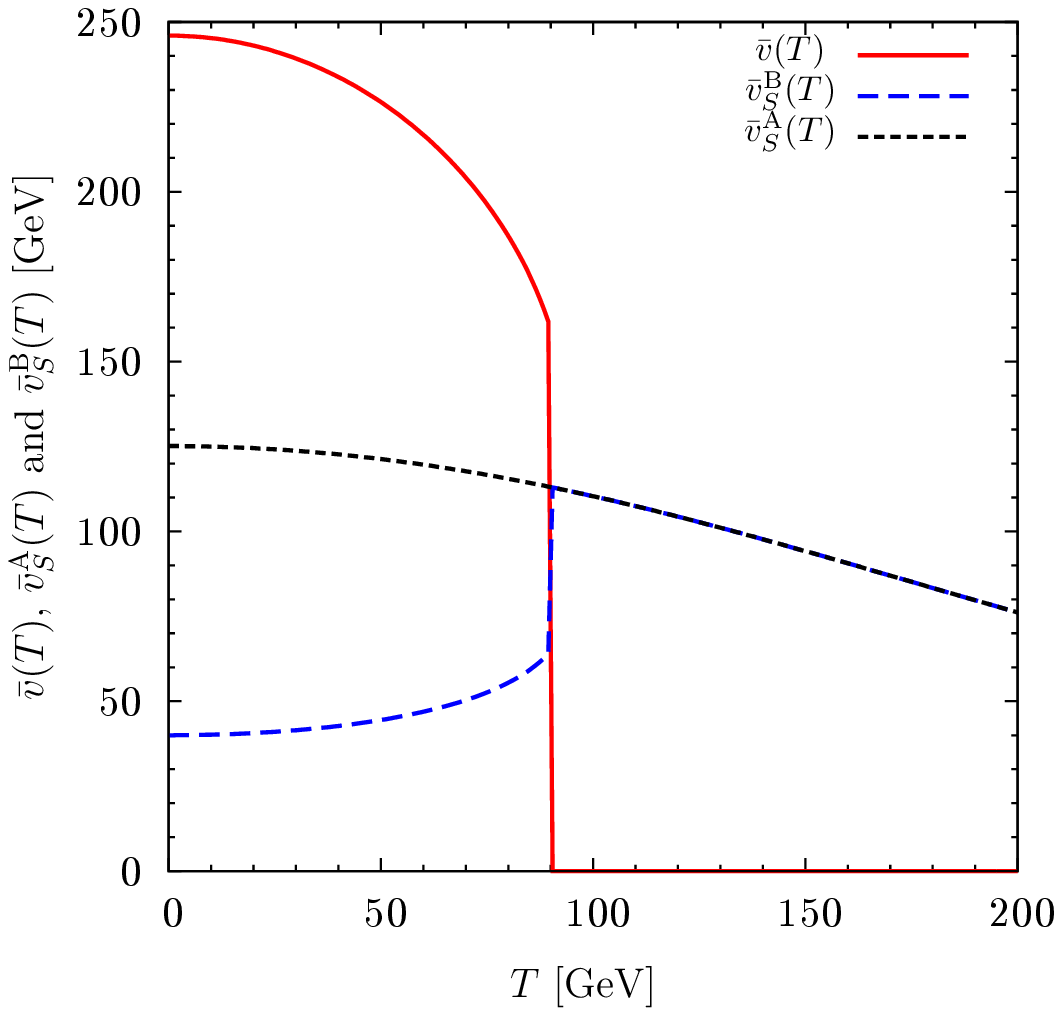}
\hspace{1cm}
\includegraphics[width=7cm]{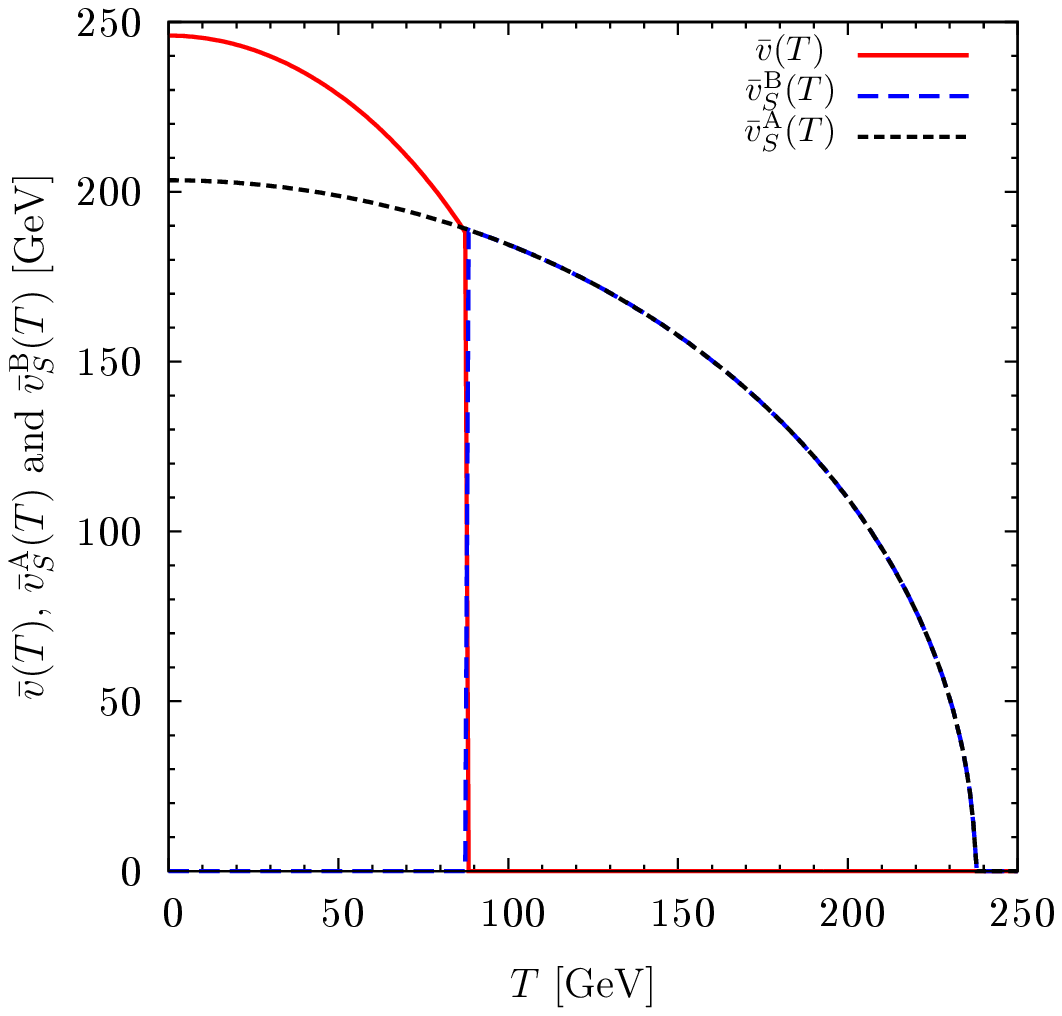}
\caption{Evolution of VEV's as a function of $T$ using the high-$T$ effective potential
in Type-(a) EWPT with $a_1\neq 0$ (Left) and Type-(c) EWPT with $a_1= 0$ (Right). 
For the former, the $A\to B$ transition is first-order, with $T_2=T_C=90.4$~GeV
and $\bar{v}(T_C) = 158.2$~GeV.
For the latter, the $O\to A$ transition is second-order while the $A\to B$ transition is first-order.
It is found that $T_1=224.6$~GeV, $T_2=T_C=99.8$~GeV and $\bar{v}(T_C) = 167.0$~GeV.
}
\label{fig:VEVs_T}
\end{figure*}

%
For $a_1\not=0$, the high-$T$ minimum will no longer lie at the origin but will be shifted by $-a_1$ along the $\varphi_S$ direction  to O$^\prime$, as illustrated in Fig.~\ref{fig:PT_nonzeroa1}.  The transition from O$^\prime$ $\to$ A is a continuous process as the temperature drops.

In what follows, we exclusively explore illustrative two scenarios: Type-(a) EWPT with $a_1\neq 0$ (corresponding to Fig.~\ref{fig:PT_nonzeroa1}) and Type-(c) EWPT with $a_1=0$.  More explicitly, we consider two scenarios:
\begin{itemize}
\item[S1:] $m_{H_2}=230$~GeV, $v_{S0}=40$~GeV, $a_1=-(110~{\rm GeV})^3$,
\item[S2:] $m_{H_2}=m_A$, $v_{S0}=0$~GeV, $\delta_2=0.55$, $d_2=0.5$,
\end{itemize}
and $m_A=m_{H_1}/2=62.5$ GeV in both cases.
\footnote{We will take $m_A$ as a variable parameter when discussing the DM phenomenology 
in Sec.~\ref{sec:DM}. Furthermore, $H_2(=S)$ is also the DM candidate in S2.}
In Fig.~\ref{fig:VEVs_T}, the evolution of the VEV's with temperature are plotted for S1 (left) and S2 (right).
For the former, there is no {O $\to$ A} transition, as shown in Fig.~\ref{fig:PT_nonzeroa1},
and we find that the { A $\to$ B} transition is first-order since $\bar{v}$
has a discontinuity at $T_2=T_C=90.4$~GeV with $\bar{v}(T_C) = 158.2$~GeV. 
For the latter, on the other hand,
one can see that the { O $\to$ A} transition is second-order, with $T_1=224.6$~GeV, while the { A $\to$ B} transition is first-order, with $T_2=T_C=99.8$~GeV, $\bar{v}(T_C) = 167.0$ GeV, $\bar{v}_{S}^{\rm B}(T_C)=0$ and $\bar{v}_{S}^{\rm A}(T_C)=168.4$~GeV.

\begin{figure*}[t]
\center
\includegraphics[width=6cm]{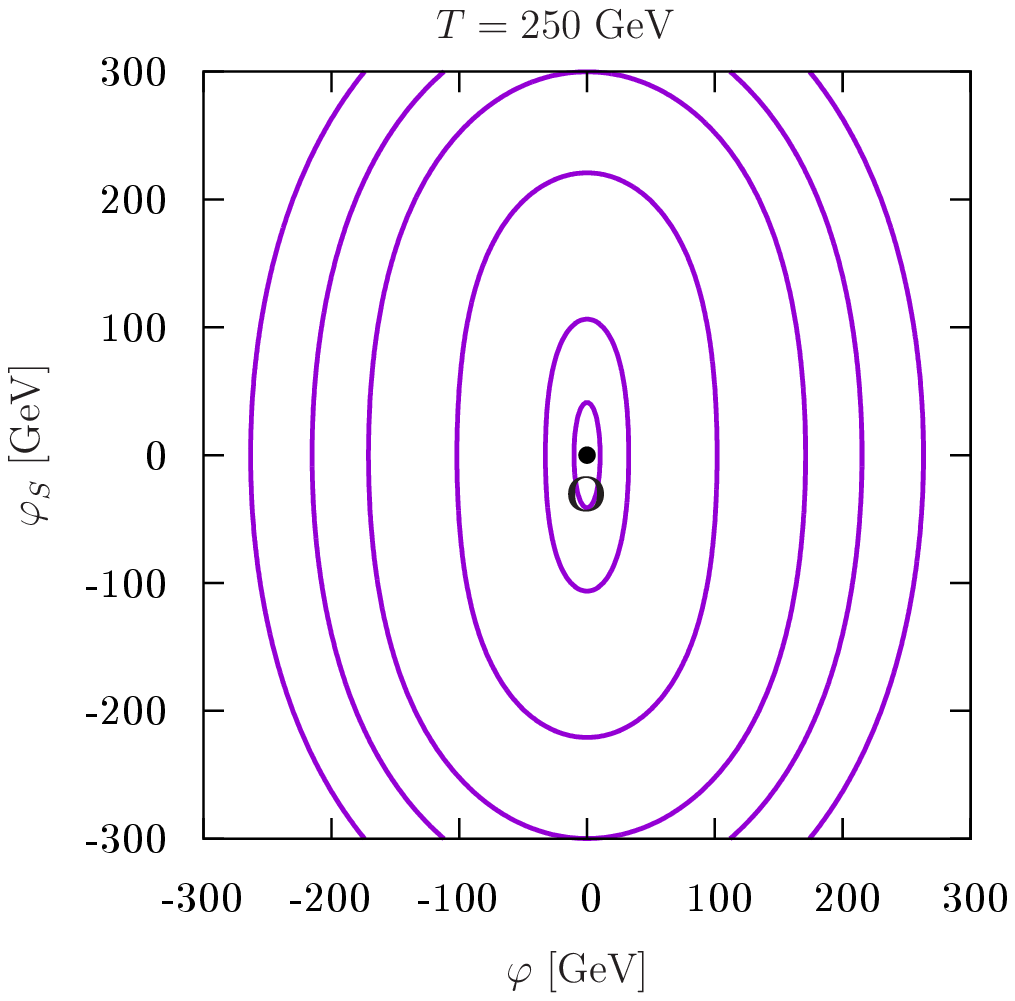}
\hspace{0.5cm}
\includegraphics[width=6cm]{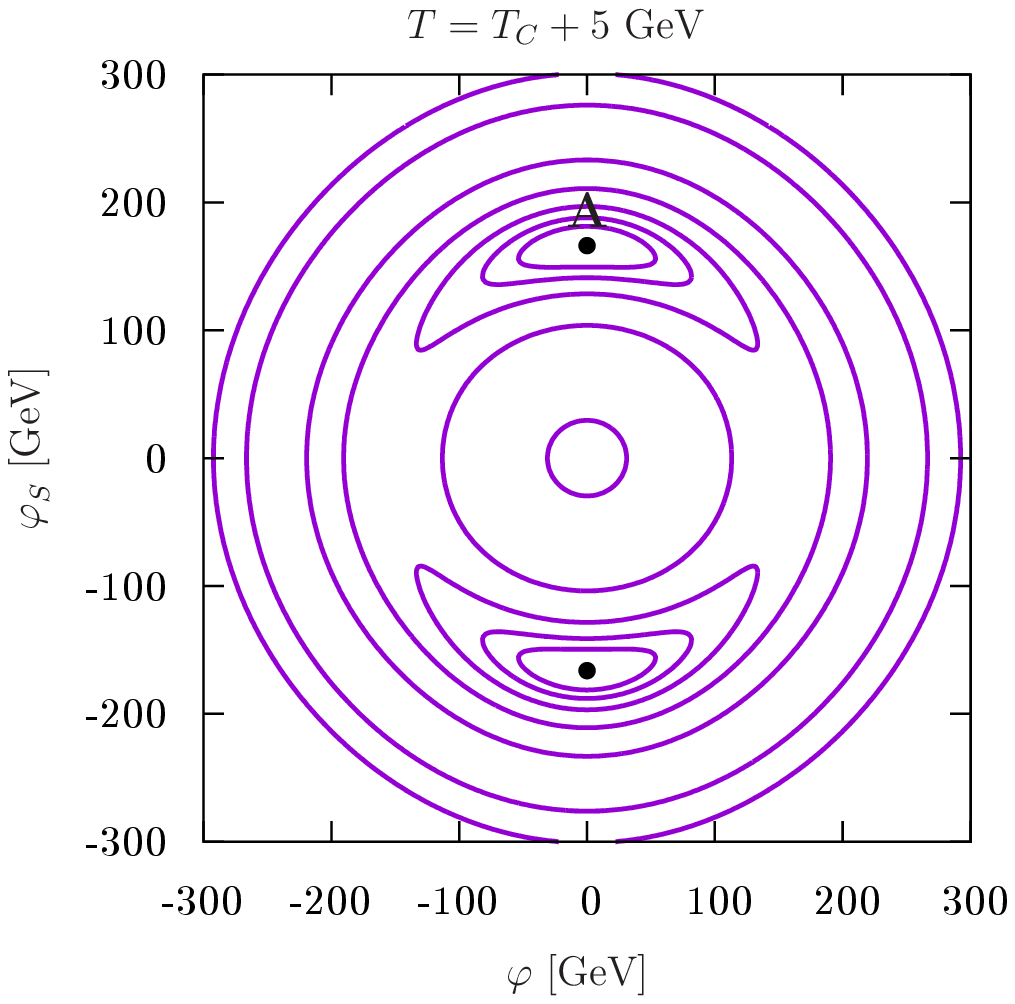}
\\[0.5cm]
\includegraphics[width=6cm]{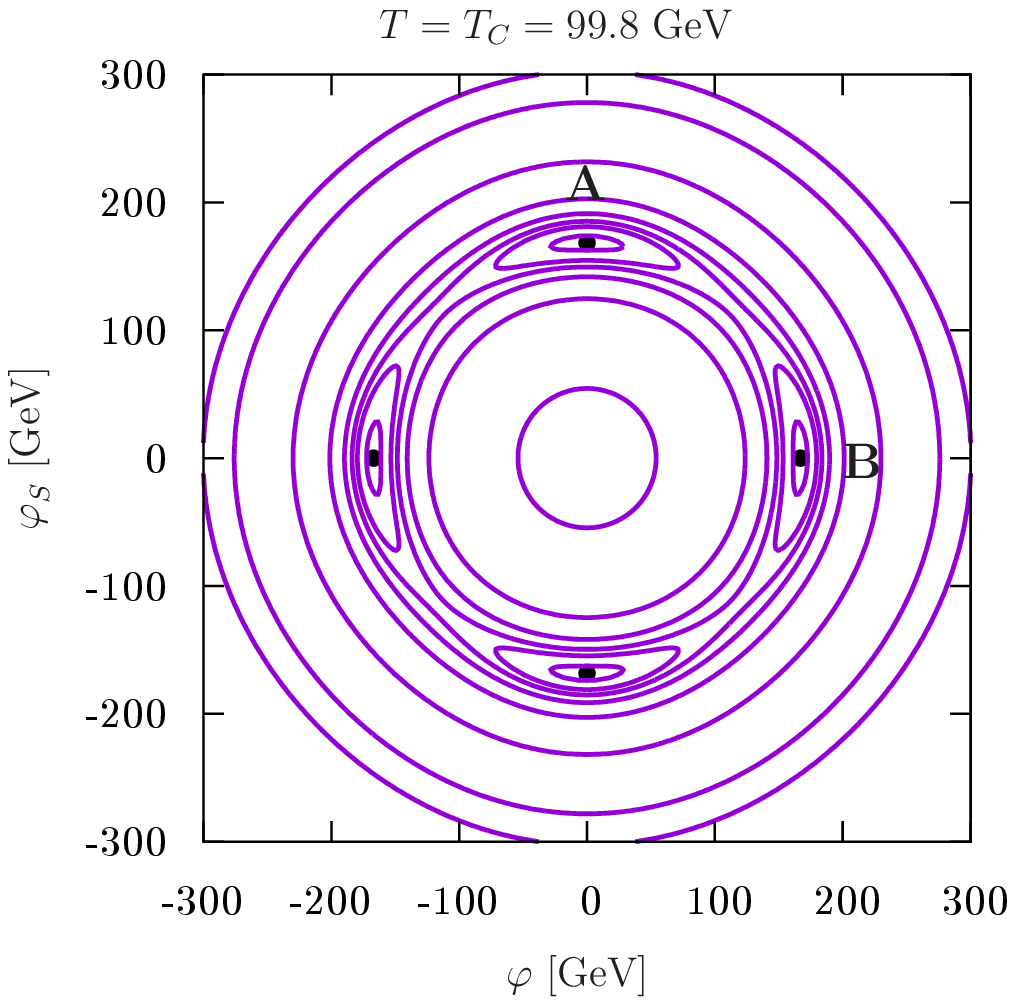}
\hspace{0.5cm}
\includegraphics[width=6cm]{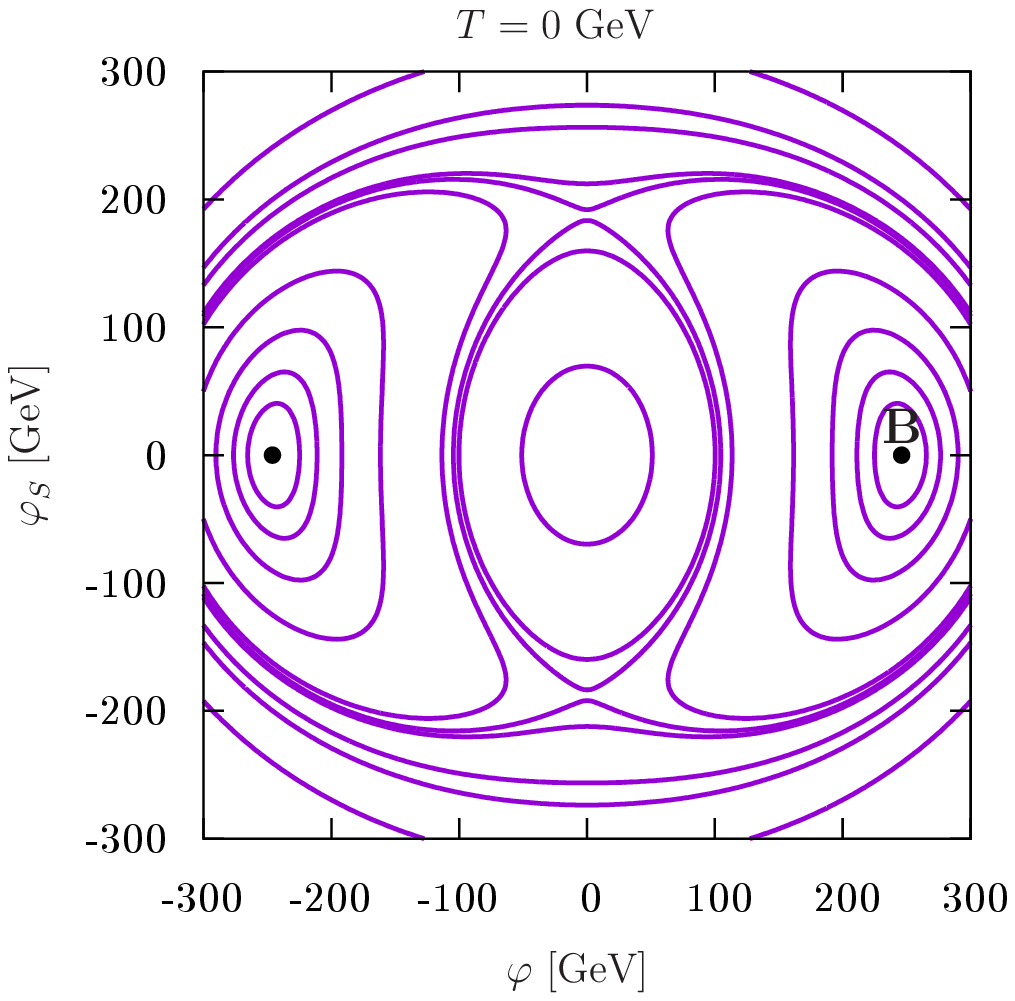}
\caption{Contours of high-$T$ effective potential at $T=250$ GeV (Upper Left), $T_C+5$ GeV (Upper Right), $T_C$ (Lower Left) and 0 GeV (Lower Right),
where $T_C=99.8$ GeV, $\bar{v}(T_C)=167.0$ GeV,
$\bar{v}_{S}^{\rm B}(T_C)=0$ GeV, $\bar{v}_{S}^{\rm A}(T_C)=168.4$ GeV in the case of S2.
The black dots denote the minima of the potential.
}
\label{fig:Veff}
\end{figure*}

Fig.~\ref{fig:Veff} shows contours of the high-$T$ effective potential 
at $T=250$~GeV (upper left), $T_C+5$~GeV (upper right), $T_C$ (lower left),  and 0~GeV (lower right) in the case of S2. The minima of the potential are indicated by the large black dots. One can see that at $T_C$ there exists a barrier between the pure singlet VEV [point A of Fig.~\ref{fig:PTs}(c)] and the pure doublet VEV  point B of Fig.~\ref{fig:PTs}(c)]. At $T=0$, the only minimum occurs along the doublet direction whereas the extremum along the singlet direction is a saddle point.

 \section{Baryogenesis}\label{sec:EWBG} 
EWBG requires that the transition to the EWSB vacuum { B} be strongly first-order, associated with both bubble nucleation and quenching of the EW sphalerons inside the bubbles. The nucleation rate, $\Gamma_N$, is governed by the three-dimensional effective action, $S_3$, which is typically computed using the bounce solution in the presence of $V_\mathrm{eff}(\boldsymbol{\varphi}; T)$ at the one-loop order. In this case, the gauge dependence must be treated with some care. In this paper, we use the high-$T$ effective potential, Eq.~(\ref{V_highT_cxSM}), for the evaluation of $\Gamma_N$ as a first step toward more complete analyses. For the regions of parameter space explored in this study, the tree-level $T=0$ potential contains a barrier between the Higgs phase vacuum and the electroweak symmetric vacuum. Consequently, we need not consider scenarios with a thermally-induced barrier that introduces problematic gauge-dependence\footnote{We also observe that for scenarios considered here, use of an $\hbar$-expansion to obtain a gauge-invariant $S_3$ is, in principle, possible.}.
 
The nucleation temperature, $T_N$, is defined as a temperature that satisfies $\Gamma_N(T_N)/H^3(T_N)=H(T_N)$ with $H(T)$ being a Hubble constant~ \cite{Linde:1981zj}.
\footnote{Since the bubble velocity is less than the speed of light, a single nucleated bubble is not able to convert the entire region of the Universe to the broken phase. Thus,  $T_N$ defined here merely represents an upper bound for the onset of the EWPT.}
With an approximation of $\Gamma_N$ discussed in Ref.~\cite{Linde:1981zj}, 
the above condition may be cast into the form
\begin{align}
&\frac{S_3(T_N)}{T_N}-\frac{3}{2}\ln\left(\frac{S_3(T_N)}{T_N}\right) \nonumber \\
&= 152.59-2\ln g_*(T_N)-4\ln\left(\frac{T_N}{100~{\rm GeV}}\right),
\label{TN}
\end{align}
where $g_*(T)$ is the number of relativistic degrees of freedom in the plasma.
Roughly speaking, Eq.~\eqref{TN} implies that $S_3(T_N)/T_N\simeq150$ is needed for the development of EWPT~\cite{Linde:1981zj}.
As we demonstrate below, choosing parameters to decrease the broken phase sphaleron rate (thereby suppressing baryon number washout) also leads to a larger
$S_3(T_N)/T_N$. Eventually, one cannot fulfill the condition of Eq.~(\ref{TN}).

Assuming a sufficient nucleation rate, preservation of the baryon asymmetry inside the bubbles imposes a requirement on the EW sphaleron transition rate: 
\begin{equation}
\Gamma_\mathrm{sph} = A_\mathrm{sph}(T) \exp[ -E_\mathrm{sph}(T)/T] \ \ \ ,
\end{equation}
where $E_\mathrm{sph}$ is the sphaleron energy, $A_\mathrm{sph}$ is a temperature-dependent pre-factor, and $T\lesssim T_N$ with $T_N$ being the nucleation temperature that typically lies just below $T_C$. The effect of baryon number washout inside the bubbles is characterized by the washout factor 
\begin{equation}
 S = \frac{n_B(\Delta t_\mathrm{EW})}{n_B(0)} ~,
\end{equation}
where $n_B(t)$ is the baryon number density at a time $t$ after the onset of the transition, $n_B(0)$ is the initial baryon number density, and $\Delta t_\mathrm{EW}$ is the duration of the EWPT. Requiring that $S> \exp (-X)$, one obtains the baryon number preservation criterion, or BNPC~\cite{Patel:2011th}:
 \begin{multline}
\label{eq:washout2}
\frac{E_\mathrm{sph}(T_C)}{T_C}-7\ln\frac{\bar{v}(T_C)}{T_C} >\\
 -\ln X-\ln\Big(\frac{\Delta t_\text{EW}}{t_H}\Big)+\ln \mathcal{QF}+\ln\kappa\, ,
\end{multline}
where $t_H$ is the Hubble time, $\kappa$ is the fluctuation determinant about the classical sphaleron solution, and $\mathcal{Q}$ and $\mathcal{F}$ encode the effects of rotational and translational zero modes as well as the unstable mode about the sphaleron\footnote{The quantity $X$ parametrizes the degree to which the initial baryon asymmetry may be diluted by sphaleron processes. Its value will depend on the initial value of the asymmetry obtained from a computation of the CPV transport dynamics in a given model.}.

It is convenient to express $E_\mathrm{sph}$ in terms of an energy scale $\Omega(T)$ associated with the EWSB that is typically of order $T_C$. To this end, we write
\begin{equation}
E_{\rm sph}(T)=\frac{4\pi \Omega(T)}{g_2}\mathcal{E}(T)~.
\end{equation}

When the only scalar fields in the theory are SU(2$)_L$ doublets, the natural choice for $\Omega(T)$ is ${\bar v}(T)$. For the cxSM, there exists a second possibility: $\sqrt{\bar{v}^2(T)+(\bar{v}_S^{\rm A}(T)-\bar{v}_S^{\rm B}(T))^2}$. Either choice is acceptable, as the BNPC depends on $E_\mathrm{sph}$ and the difference in the choice of $\Omega(T)$ will be compensated by the corresponding convention for $\mathcal{E}(T)$.
Here we follow Refs.~\cite{Funakubo:2005bu,Ahriche:2007jp} where it is argued that $\Omega(T)={\bar v}(T)$ encapsulates the primary $T$-dependence of the sphaleron energy. Nevertheless, we find that the residual $T$-dependence of $\mathcal{E}$ can be non-negligible in some cases.
The detailed calculation of $E_{\rm sph}(T)$ is given in Appendix~\ref{app:Esph}.

From these considerations, one obtains from the BNPC (\ref{eq:washout2}) a requirement on the ratio ${\bar v}(T_C)/T_C$:
\begin{align}
\frac{{\bar v}(T_C)}{T_C} \gtrsim \zeta_{\rm sph}(T_C) ~.
\label{sph_dec}
\end{align}
In the literature, one often finds this requirement quoted as $v(T_C)/T_C \gtrsim 1$, where $v(T_C)$ is computed using the full one-loop effective potential $V_\mathrm{eff}$ rather than $V^{{\rm high\mathchar`-}T}$~\cite{Funakubo:2009eg,Fuyuto:2014yia,Ahriche:2014jna,Fuyuto:2015vna}. As discussed in Ref.~\cite{Patel:2011th}, this procedure, as well as the conventional method for computing $T_C$, introduces an unphysical gauge dependence. 
In what follows, we will perform a gauge-invariant computation.
We also address the impact of the $\mu$-dependence by implementing a RG-improved analysis. These and other theoretical issues associated with the BNPC and $\Gamma_N$ are discussed below. 


\section{Gauge-invariant method beyond the leading order}
\label{sec:girge}
Here, we delineate the gauge-invariant treatment for EWPT and sphaleron rate.
Determination of $T_C$ and $\bar{v}_C$ using the high-$T$ effective potential is obviously gauge independent. 
Beyond this order, however, the potential barrier inherently depends on the gauge fixing parameter,
which may lead to the gauge-dependent $T_C$ and $v_C$ as in the ordinary method.
Nevertheless, the gauge-invariant $T_C$ can still be obtained 
by use of a method advocated in Ref.~\cite{Patel:2011th} (PRM scheme).
Here we will outline the method briefly.

The Nielsen-Fukuda-Kugo identity~\cite{Nielsen:1975fs,Fukuda:1975di} states that energies at station points of the effective potential are free from the gauge fixing parameter,
which is described by 
\begin{align}
\frac{\partial V_{\text{eff}}(\varphi)}{\partial \xi}  = -C(\varphi,\xi)\frac{\partial V_{\text{eff}}(\varphi)}{\partial \varphi},
\end{align}
where $C(\varphi, \xi)$ denotes some functional (for the explicit form, see Ref.~\cite{Nielsen:1975fs}).
We determine $T_C$ and $v(T_C)$ in such a way that the above identity is satisfied order by order in the perturbative expansion. 
Let us expand $V_{\text{eff}}$ and $C$ in powers of $\hbar$ as
\begin{align}
V_{\text{eff}}(\varphi) &= V_0(\varphi)+\hbar V_1(\varphi)+\hbar^2 V_2(\varphi)+\cdots, \\
C(\varphi,\xi) &= c_0+\hbar c_1(\varphi)+\hbar^2 c_2(\varphi)+\cdots.
\end{align}
For example, the identity to $\mathcal{O}(\hbar)$ is found to be
\begin{align}
\frac{\partial V_1}{\partial \xi} &= -c_1\frac{\partial V_0}{\partial\varphi}.
\end{align}
Note that the $\xi$ dependence of $V_1$ drops out at the points where the {\it tree-level} potential is extremized, which differs from the extremum 
of $V_1$.
 
Correspondingly, $T_C$ to $\mathcal{O}(\hbar)$ is determined by the following degeneracy condition

\begin{align}
V_0(\boldsymbol{v}_0^{\rm A})+V_1(\boldsymbol{v}_0^{\rm A}; T_C)
=V_0(\boldsymbol{v}_0^{\rm B})+V_1(\boldsymbol{v}_0^{\rm B}; T_C) ~,
\label{def_Tc}
\end{align}
where $\boldsymbol{v}_0^{\rm A}$ denote the VEV's of phase A while $\boldsymbol{v}_0^{\rm B}$ those of phase B as defined above.  

On the other hand, $\bar{v}(T_C)$ in the PRM scheme is determined by the high temperature potential $V^{\text{high-}T}$ given in Eq.~(\ref{V_highT_cxSM}), which is manifestly gauge invariant.

Though the so-called ring diagrams can also be implemented in a gauge invariant manner, knowledge of $V_2$ is indispensable
for the consistency of the calculation.   
Since $V_2$ in this model is not available to date, we will confine ourselves to the $\mathcal{O}(\hbar)$ calculation in this paper.

The appearance of the renormalization scale $\mu$ in $V_{\rm CW}$ in Eq.~(\ref{V1}) can lead to a significant $\mu$-dependence for  
$T_C$. To alleviate this dependence, we will replace  $V_0(\boldsymbol{\varphi})$ appearing in the degeneracy condition (\ref{def_Tc}) by the RG-improved potential $\widetilde{V}_0(\boldsymbol{\varphi})$. 
More explicitly, we replace all the parameters in $V_0$ with the running ones 
\begin{align}
\lefteqn{\widetilde{V}_0(\varphi, \varphi_S)}\nonumber\\
&= \frac{m^2(\mu^2)}{4}\varphi^2+\frac{\lambda(\mu^2)}{16}\varphi^4
	+\frac{\delta_2(\mu^2)}{8}\varphi^2\varphi_S^2	
	+\frac{b_2(\mu^2)}{4}\varphi_S^2\nonumber\\
&\quad+\frac{d_2(\mu^2)}{16}\varphi_S^4 
+\sqrt{2}a_1(\mu^2)\varphi_S+\frac{b_1(\mu^2)}{4}\varphi_S^2.
\end{align}
Here, we use the one-loop $\beta$ functions~\cite{Gonderinger:2012rd} to evaluate the running parameters, and the RG effects on $\varphi$ and $\varphi_S$ are ignored as they are negligible. 
Note that the other parameters appearing in Eq.~(\ref{def_Tc}) remain unchanged in order not to spoil the gauge independence to this order.
Similarly, $V_0(\boldsymbol{\varphi})$ in Eq.~(\ref{V_highT_cxSM}) remains as is since the renormalization scale does not enter the high-$T$ effective potential.  In what follows, we numerically demonstrate that this procedure eliminates the otherwise problematic $\mu$-dependence while maintaining gauge invariance.


\section{Numerical Analysis}
\label{sec:numerics}

In this section, we perform a numerical analysis taking theoretical and experimental constraints
into account.
The  bounded-from-below conditions for the scalar potential are imposed as
\begin{align}
\lambda >0,\quad d_2>0,\quad -\sqrt{\lambda d_2}<\delta_2,
\end{align}
where the last condition is  only applied to the $\delta_2<0$ case.
Furthermore, the absolute values of the quartic couplings are restricted to be less than $4\pi$
as a simple criterion of perturbativity (for more detailed analysis and some subtleties, 
see Ref.~\cite{Gonderinger:2012rd}). 

The quantities $m_{H_2}$ and $\alpha$ are constrained by direct searches for a heavy Higgs boson at the LHC, measurements of Higgs signal strengths~\cite{Profumo:2014opa,Costa:2014qga},
and electroweak precision observables.
For example, utilizing the two gauge boson decay modes, $m_{H_2}$
is bounded as a function of $\sin^2\alpha$ with $\mathcal{B}_{\rm new}=0.0$, 0.2 and 0.5, 
where $\mathcal{B}_{\rm new}$ denotes the non-SM contribution to decay width~\cite{Khachatryan:2015cwa}.
For the mixing angle, it is found that 
$\cos\alpha\gtrsim 0.8$ for $m_{H_2}\gtrsim 250$ GeV to be consistent with the EW observables
within the 3$\sigma$ level~\cite{Costa:2014qga}.

\subsection{S1 Case}
\begin{figure}[t]
\center
\includegraphics[width=7cm]{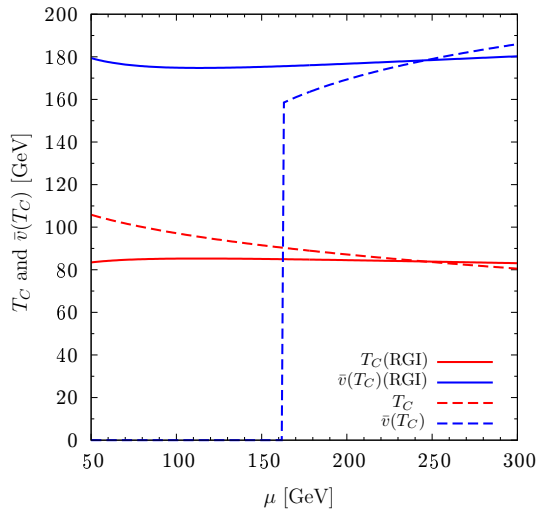}
\caption{Renormalization scale dependence { of} $T_C$ and $\bar{v}_C$ in S1.
The { dashed} curves are calculated based on the original PRM scheme to $\mathcal{O}(\hbar)$ 
while the solid ones are the RG-improved version.}
\label{fig:EWPT_mudep}
\end{figure}
 Recall that scenario S1 follows the two-step history of Fig.~\ref{fig:PT_nonzeroa1} with the high-$T$ minimum lying away from the origin due to the non-vanishing $a_1$.
Fig.~\ref{fig:EWPT_mudep} shows the $\mu$-dependence of $T_C$ and the corresponding $\bar{v}(T_C)$ with and without the RG improvement.  
The input parameters are the same as in the left plot of Fig~\ref{fig:VEVs_T}.
The solid curves represent the former and the dashed ones the latter.  One can see that $T_C$ has a some renormalization scale dependence before the RG improvement. This results in uncertainties in the determination of $\bar{v}(T_C)$, which is shown in the plot to vary from zero to nonzero.  This makes it difficult to reach a conclusion whether or not the EWPT is of first order.
After the RG improvement, however, the renormalization scale dependence is substantially 
alleviated, and the EWPT is seen to be strongly first-order.

\begin{figure}
\center
\includegraphics[width=7cm]{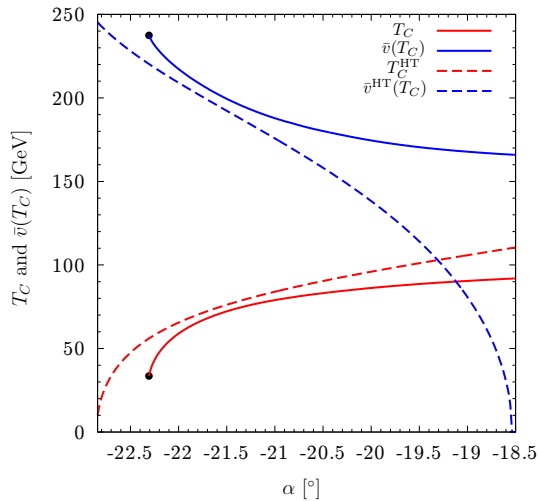}
\caption{$T_C$ and $\bar{v}(T_C)$ as functions of $\alpha$ in S1. 
The solid curves are obtained by the PRM scheme with RG improvement while the dashed ones by the high-$T$ effective potential in Eq.~(\ref{V_highT_cxSM}). }
\label{fig:EWPT_alpha_gl}
\end{figure}
 
As discussed above, since the  mixing angle $\alpha$ 
is one of the most important parameters for both LHC phenomenology and for a SFOEWPT, we first quantify the
 $\alpha$-dependence of $T_C$ and $\bar{v}$.
In Fig.~\ref{fig:EWPT_alpha_gl}, $T_C$ and $\bar{v}(T_C)$ are plotted as functions of $\alpha$  
where the solid curves correspond to the PRM scheme with RG improvement, while the dashed ones the are calculated by use of the high-$T$ potential (HT calculation). One can see that $T_C$ decreases (more favorable for SFOEWPT) as $\alpha$ decreases in both cases.
Recall that a positive, increasing $\delta_2$ also leads to a decreasing $T_C$, as  seen from the analytic formula Eq.~(\ref{Tc}) (with $m^2<0$).
For $\alpha\lesssim -22.8^\circ$, however, phase $A$ becomes the global minimum,
yielding an upper bound $\delta_2\lesssim 2.7$ in the HT calculation.
In the PRM calculation, there is an end point around $\alpha\simeq -22.3^\circ$ indicated by the black dots.
Below this point, the vacuum energy of  phase B is higher than that of phase $A$,
even at $T=0$.  Consequently, the degeneracy condition~(\ref{def_Tc}) is never satisfied.

Note that $T_C<T_C^{\rm HT}$ in all the range 
and that the differences between the HT and PRM results become more prominent as $\alpha$ increases.
Moreover, the tree-level potential barrier  disappears for $\alpha\gtrsim -18.5^\circ$. Thus, in the HT calculation, the EWPT transition is second order in this region, and   
$\bar{v}$ can correspondingly  be cast into the form
$\sqrt{1-(T/T_0)^2}$, where $T_0$ is the critical temperature of the second-order EWPT.
In the PRM calculation, on the other hand, a small loop-induced contribution to the potential  remains, and one finds that 
$T_C<T_0$. Since this leads to $\bar{v}(T_C)>0$,
one might na\"ively interpret this to indicate the presence of a SFOEWPT.
However, phase A in this case becomes a saddle point rather than a local minimum, 
and thus the region $\alpha \agt -18.5^\circ$ is, in fact, excluded.

As emphasized in Ref.~\cite{Patel:2011th}, the $\mathcal{O}(\hbar)$ computation of $T_C$ 
is likely an underestimate. In the SM, inclusion of higher order contributions, either in perturbation theory or lattice computations, yields a larger value of $T_C$. The corresponding theoretical uncertainty can be significant. 
Even though a quantitatively robust statement awaits a more precise analysis,  we are able to identify some general trends:
 the realization of SFOEWPT in this model needs 
a negative $\alpha$ (corresponding to positive $\delta_2$) but not 
 too large in magnitude, as shown by
the range of Fig.~\ref{fig:EWPT_alpha_gl}.

\begin{figure}[t]
\center
\includegraphics[width=7cm]{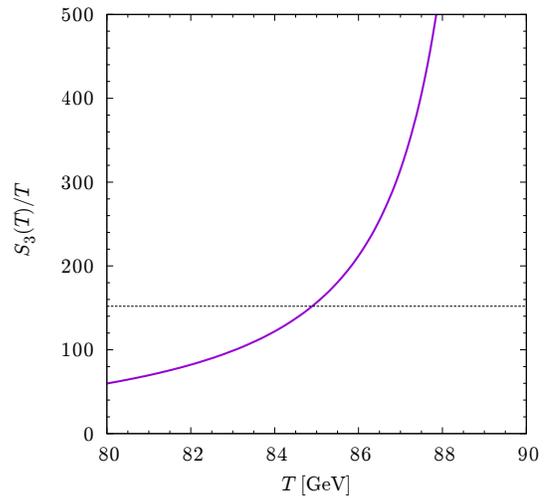}
\caption{$S_3(T)/T$ vs. $T$ for $\alpha=-20.5^\circ$ in S1. $S_3(T)$ is evaluated by use of the high-$T$ effective potential in Eq.~(\ref{V_highT_cxSM}). The dotted horizontal line satisfies the condition in Eq.~(\ref{TN}). 
In this case, we have $T_C^{\rm HT}=90.4$~GeV and $T_N=84.9$~GeV, with the latter closer to $T_C$
calculated in the PRM scheme.}
\label{fig:S3overT_T}
\end{figure}
As discussed in Section~\ref{sec:EWBG}, the actual beginning of the EWPT occurs at a temperature $T_N$ that is somewhat below the temperature at which the effective potential has two degenerate minima.~\footnote{Such a phenomenon that the phase transition delays its occurrence until the nucleation temperature lower than the critical temperature is called supercooling.}
If $T_N$ is sufficiently close to $T_C$, we can approximate $T_C$ 
as the transition temperature, and thus $E_{\rm sph}(T_C)/T_C\simeq E_{\rm sph}(T_N)/T_N$ approximately holds.
It is useful to characterize the degree of supercooling by the quantity
\begin{equation}
\label{eq:super}
\Delta \equiv \frac{T_C-T_N}{T_C}.
\end{equation}
It is known that $\Delta=\mathcal{O}(0.1)\%$ 
in the minimal supersymmetric SM (MSSM) case (see, {\it e.g.}, Refs.~\cite{Moreno:1998bq,Funakubo:2009eg}). 

Here, we calculate $S_3(T)$ to find $T_N$ using the high-$T$ effective potential in Eq.~(\ref{V_highT_cxSM}).
In Fig.~\ref{fig:S3overT_T}, the solid curve shows $S_3(T)/T$ as a function of $T$ for $\alpha=-20.5^\circ$ in S1.
The dotted line satisfies the condition in Eq.~(\ref{TN}), from which we obtain $S_3(T_N)/T_N=152.01$ and 
$T_N=84.9$~GeV. Since $T_C^{\rm HT}=90.4$~GeV, one obtains $\Delta_\mathrm{HT} = 6.1\%$.

We also find that the supercooling becomes larger as $\alpha$ decreases,
and eventually the condition of Eq.~(\ref{TN}) cannot be fulfilled for $\alpha\lesssim-21.4^\circ$,
rendering a more stringent lower bound on $\alpha$ than the vacuum degeneracy condition mentioned above.
For the critical $\alpha=-21.4^\circ $, we obtain $T_C^{\rm HT}=78.1$~GeV and $T_N=47.3$~GeV,
leading to $\Delta_\mathrm{HT} \simeq 39.4\%$. 

The degree of supercooling will affect the dynamics of the EWPT and the feasibility of EWBG.
As discussed above, the first-order EWPT proceeds via the bubble nucleation and expansion.
For EWBG to be successful, most of  the region in the symmetric phase has to be 
converted into the broken phase via the bubble expansion rather than the nucleation
since the baryon asymmetry is generated by the scatterings of the particles
with the growing bubbles.
If the supercooling $\Delta$ is large, however, the EWPT may proceed mostly via the nucleation
since the nucleation rate $\Gamma_N$ gets enhanced, spoiling the EWBG mechanism.
Prior to the discovery of the Higgs boson and top quark, these dynamics of the EWPT in the SM were studied in Ref.~\cite{Carrington:1993ng}.
For $m_h=60$ GeV and $m_t=120$ GeV, the supercooling is found to be around 0.25\% . 
About 10\% of the symmetric-phase region is converted into
the broken phase when  bubble nucleation completes, and the rest of the conversion results from
the bubble expansion. An analogous study is needed for cxSM in order to assess the viability of EWBG.

\begin{figure}
\center
\includegraphics[width=7cm]{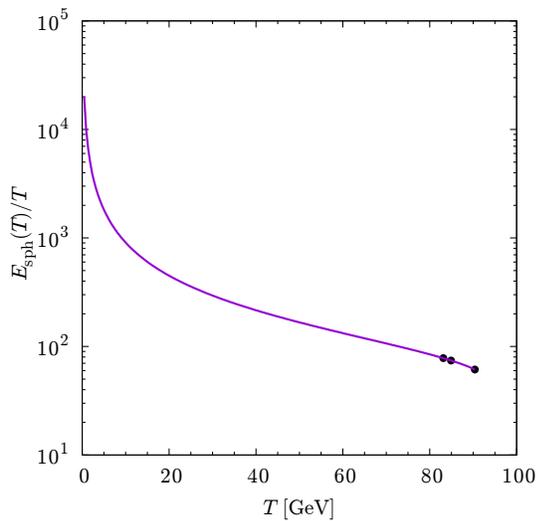}
\caption{$E_{\rm sph}(T)/T$ as a function of $T$,
where $E_{\rm sph}(T)$ is calculated using the high-$T$ effective potential in Eq.~(\ref{V_highT_cxSM}).
From right to left, the dots mark for $E_{\rm sph}(T_C^{\rm HT})/T_C^{\rm HT}=61.31$, 
$E_{\rm sph}(T_N)/T_N=74.23$ and $E_{\rm sph}(T_C)/T_C=78.00$,
where $T_C^{\rm HT}=90.4$~GeV, $T_N=84.9$~GeV and $T_C=83.1$~GeV.}
\label{fig:EsphoverT_T}
\end{figure}
We now turn to the final set of questions posed in Section \ref{sec:intro}: To what extent does ${\bar v}(T)$ characterize the sphaleron energy, 
$E_{\rm sph}(T)$? To that end, we first plot in
Fig.~\ref{fig:EsphoverT_T} the ratio $E_{\rm sph}(T)/T$ vs. $T$ for $\alpha=-20.5^\circ$ in S1, where $E_{\rm sph}(T)$ is estimated based on the high-$T$ effective potential in Eq.~(\ref{V_highT_cxSM}).
From right to left, the three dots mark the results for $E_{\rm sph}(T_C^{\rm HT})/T_C^{\rm HT}=61.31$, 
$E_{\rm sph}(T_N)/T_N=74.23$ and $E_{\rm sph}(T_C)/T_C=78.00$ using the values of $T_C^{\rm HT}$, $T_N$ and $T_C$ given above. 
Recall that for this value of $\alpha$, the tree-level potential admits a barrier between the phases A and B, so that the EWPT is first order, even in the HT framework. Thus, $T_C^{\rm HT} < T_0$, where $T_0$ is the temperature at which the gauge invariant scale ${\bar v}(T)$ vanishes, implying a vanishing $E_{\rm sph}(T)$ in the computational framework adopted here. 
The endpoint at $T_C^{\rm HT}$ simply implies that of the computational approaches discussed here, the one giving the maximum $T_C$, and thus, the minimum (but non-vanishing)  $E_{\rm sph}(T_C)/T_C$, is the HT computation. 

\begin{figure}
\center
\includegraphics[width=7cm]{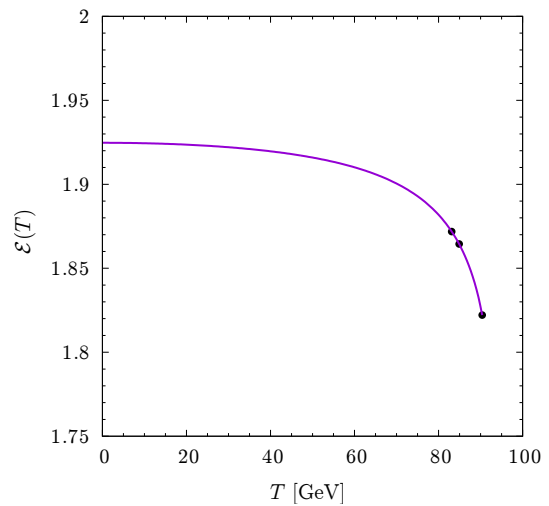}
\caption{$\mathcal{E}(T)=g_2E_{\rm sph}(T)/(4\pi\bar{v})$ as a function of $T$.
The three dots correspond to $\mathcal{E}(T_C^{\rm HT})=1.82$, $\mathcal{E}(T_N)=1.86$ and $\mathcal{E}(T_C)=1.87$ 
from right to left.}
\label{fig:rEsph_T}
\end{figure}

In Fig.~\ref{fig:rEsph_T}, the dimensionless sphaleron energy $\mathcal{E}(T)$ is plotted as a function of $T$.  Apparently, $\mathcal{E}(T)$ decreases as $T$ increases,
showing that the temperature dependence of $E_{\rm sph}(T)$ is not fully embodied in $\Omega(T)$, where we have taken $\Omega(T)={\bar v}(T)$ as indicated above. We conclude that 
the na{\" i}ve scaling formula $E_{\rm sph}(T)=E_{\rm sph}(0){\bar v(T)}/v_0$ is no longer valid, especially when $T$ approaches $T_C$ (for earlier studies, see Refs.~\cite{Moreno:1996zm,Funakubo:2009eg,Fuyuto:2014yia}).  
As in Fig.~\ref{fig:EsphoverT_T}, the three dots correspond to 
$\mathcal{E}(T_C^{\rm HT})=1.82$, $\mathcal{E}(T_N)=1.86$ and $\mathcal{E}(T_C)=1.87$ 
from right to left. Because the deviation from a linear dependence of $E_{\rm sph}(T)$ on ${\bar v}(T)$ is particularly pronounced in the vicinity of the critical temperature, one should apply a fair degree of caution before drawing sharp conclusions about the viability of EWBG based on a one-loop perturbative treatment of the EWPT. As we have emphasized earlier, inclusion of higher-order contributions can lead to significant variations { in} $T_C$, implying corresponding variations in the $E_{\rm sph}(T)/T$ and, thus, the efficacy of baryon number preservation in the broken phase.

\subsection{S2 Case}
This scenario corresponds to the thermal history of Fig.~\ref{fig:PTs}(c). Because the singlet VEV vanishes at $T=0$, one has $\alpha=0$. Moreover,  there is no constraint among the parameters $b_1$, $b_2$, $\delta_2$ and $d_2$
from the tadpole condition of Eq.~(\ref{tree-tad_hS}), in contrast to the S1 case. Consequently, it will be most instructive to consider the dependence of EWPT properties on the portal coupling $\delta_2$. 
For purposes of the following discussion, we also note that a necessary condition for the
existence of the nontrivial vacuum phase A in $\varphi_S$ is
\begin{align}
\left[ \bar{v}_S^{\rm A}(T) \right]^2 = -\frac{2}{d_2}\big(b_1+b_2+2\Sigma_ST^2\big)>0,
\label{vStilde}
\end{align}
which implies that $b_1+b_2$ must be negative as long as $\Sigma_S$ is positive.
While necessary,  this condition is not sufficient since a saddle point is also possible.

\begin{figure}[t]
\center
\includegraphics[width=7cm]{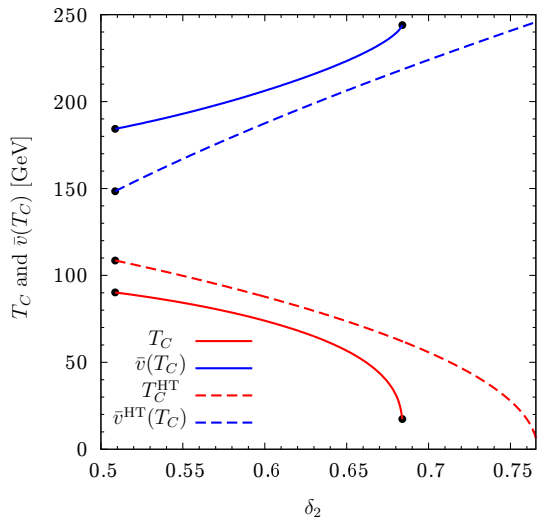}
\caption{$T_C$ and $\bar v(T_C)$ as functions of $\delta_2$.
Below the left end point ($\delta_2\simeq 0.51$) 
the EWPT is second-order  for both HT and PRM cases, while $T_C$ cannot be defined above the right end point ($\delta_2\simeq 0.68$) in the PRM case. 
For $\delta_2\gtrsim0.77$, phase A turns into the global minimum at $T=0$ in the HT case.}
\label{fig:EWPT_del2}
\end{figure}

Fig.~\ref{fig:EWPT_del2} shows the dependence of $T_C$ and  ${\bar v}(T_C)$ on $\delta_2$.
The line and color styles are the same as in Fig.~\ref{fig:EWPT_alpha_gl}.
Below the left end point at $\delta_2\simeq 0.51$, 
the potential barrier is  eliminated and thus the EWPT is of second-order.
In the HT calculation, phase A becomes the global minimum at $T=0$ for $\delta_2\gtrsim0.77$.
In this case, the EWPT never occurs.
In the PRM case, $T_C$ is lower than the HT case as explained above.
Moreover, as in the case of S1, there is a point where $T_C$ cannot be defined due to the absence of vacuum degeneracy,
which corresponds to the right end point at $\delta_2\simeq0.68$.
Our findings show that the region that is consistent with SFOEWPT is generally more limited in the PRM case.

\begin{table}[t]
\center
\begin{tabular}{|c|c|c|c|}
\hline
 $T$ [GeV] & $T_C=84.3$ & $T_C^{\rm HT}=99.8$ & $T_N=96.6$ \\ \hline
$\bar{v}(T)$ [GeV] & 193.0 & 167.0 & 173.5 \\ 
$E_{\rm sph}(T)/T$ & 84.36 & 61.67 & 66.02 \\
$\mathcal{E}(T)$ & 1.92 & 1.92 & 1.92 \\
\hline
\end{tabular}
\caption{VEV's and sphaleron energies at different temperatures, 
$T_C$, $T_C^{\rm HT}$ and $T_N$, for $\delta_2=0.55$ in S2. 
The last two columns are calculated by use of the high-$T$ effective potential (\ref{V_highT_cxSM}).}
\label{tab:S2}
\end{table}

As in the S1 case, we also evaluate $S_3(T)$, $E_{\rm sph}(T)$ and $\mathcal{E}$, 
fixing $\delta_2=0.55$.
The results are summarized in Table \ref{tab:S2}.
One can see that as in the S1 case, the degree of the supercooling is $\Delta_{\text{HT}} \simeq 3.2\%$, about one order of magnitude larger than 
the typical MSSM value~\cite{Moreno:1998bq,Funakubo:2009eg}.
However, one distinctive feature of S2 is that $\mathcal{E}$ is independent of $T$,
implying that $E_{\rm sph}(T)=E_{\rm sph}(0)\bar{v}(T)/v_0$ as in the SM.
This is due to the fact that there is no singlet Higgs contribution to $E_{\rm sph}$
since $\bar{v}_S=0$.

Before closing this section, we  comment on the impact of the DM mass  on the viability of a SFOEWPT. 
In both the S1 and S2 cases, the $T=0$ vacuum energy of phase $B$  increases
as the DM mass increases and surpasses the vacuum energy of phase $A$ at some particular value.
Hence, no transition to the phase B occurs and $T_C$ cannot be defined in the PRM calculation.
As will be discussed in the next section, a relatively large DM mass is generally required to obtain
the correct DM relic abundance and evade the LUX constraints,  but this requirement may be in conflict with the realization of a SFOEWPT except for a finely-tuned region in the vicinity of the Higgs pole.

\section{Dark Matter}\label{sec:DM}

In the cxSM, there are two possibilities for dark matter: (a) a single-component scenario, wherein the pseudoscalar particle $A$ is the dark matter particle and (b) a two-component scenario involving both $A$ and  $S$. Scenario (b) is only possible
if $v_{S0}=0$ such as the S2 case.
We will study the basic dark matter properties in  both the S1 and S2 cases\footnote{Note that $H_2=S$ in S2.}. 
We use micrOMEGAs~\cite{Belanger:2006is,Belanger:2013oya} to calculate the relic density of $A$, $\Omega_A$, and its spin-independent scattering cross section with the nucleon $N$, $\sigma_{\rm SI}^N$. To be consistent with observation, the parameters of the cxSM must not yield a relic density larger than the experimental value~\cite{Olive:2016xmw}
\begin{align}
\Omega_{\rm DM}h^2 = 0.1186\pm 0.0020 ~, \label{DM_obs}
\end{align}
nor should the rate for spin-independent scattering from the nucleon exceed the current LUX limit~\cite{Akerib:2016vxi}. For a DM mass of around 50~GeV, the maximum spin-independent cross section is $\sigma_{\rm SI}^N\simeq 1.1\times 10^{-46}~{\rm cm}^2$, assuming the DM candidate saturates the relic density.~
\footnote{Recently, the ZENON1T experiment has updated the upper bound on $\sigma_{\rm SI}^N$~\cite{Aprile:2017iyp},
which is  slightly below the LUX bound. Our conclusions are not substantially affected by the improved limit.
}

In the present case, if the relic abundance of $A$ (understood to imply the combined $A$ and $S$ abundance in S2) is less than the observed DM abundance, $\sigma_{\rm SI}^N$ should be scaled as
\begin{align}
\tilde{\sigma}_{\rm SI}^N = \sigma_{\rm SI}^N\left(\frac{\Omega_A}{\Omega_{\rm DM}}\right) ~.
\label{tildeSigmaSI}
\end{align}
For the maximum allowed $\Omega_Ah^2$ we will use the central value of Eq.~(\ref{DM_obs}).

\begin{figure*}
\center
\includegraphics[width=7cm]{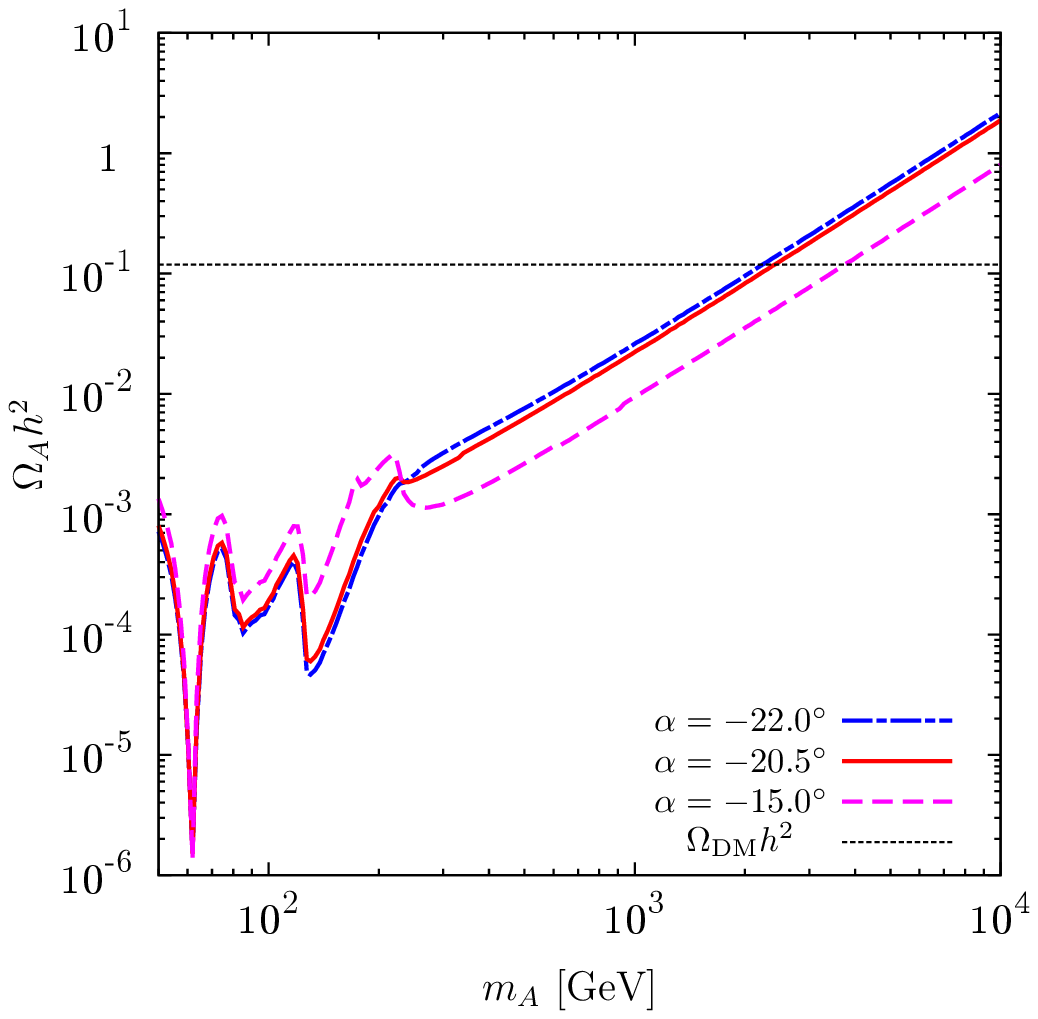}
\hspace{1cm}
\includegraphics[width=7.1cm]{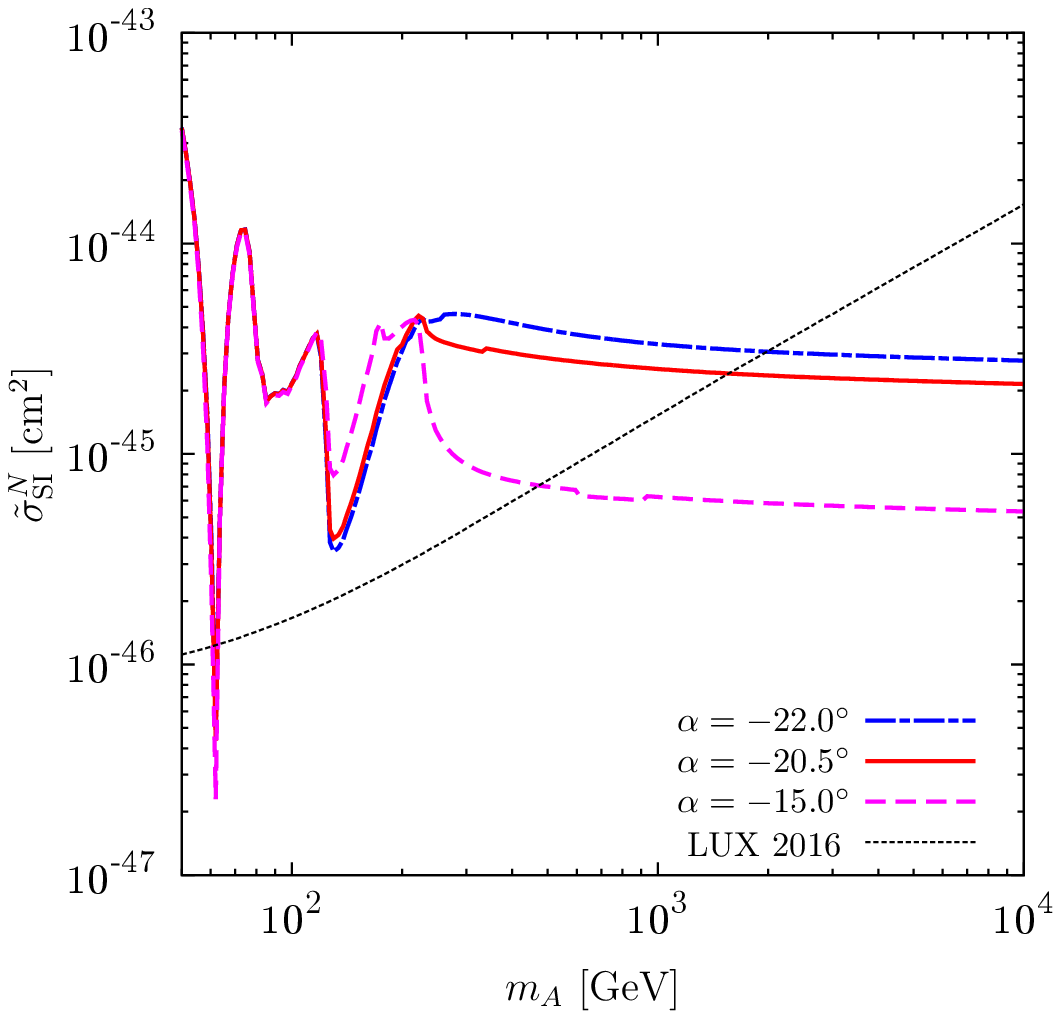}
\caption{Scalar dark matter for case S1: (Left) Relic density of $A$ with $\alpha=-22.0^\circ$, $-20.5^\circ$ and $-15.0^\circ$. 
(Right) Scaled spin-independent DM-nucleon scattering cross section.
The allowed region is only at $m_A\simeq m_{H_1}/2\simeq 62.5$ GeV if a SFOEWPT is required.
}
\label{fig:DM_nonInert}
\end{figure*}
In the left plot of Fig.~\ref{fig:DM_nonInert}, $\Omega_A h^2$ is plotted as a function of $m_A$ for case S1.
We take $\alpha=-22.0^\circ$ (blue, dot-dashed), $20.5^\circ$ (red, solid) and $-15.0^\circ$ (magenta, dashed). $\Omega_{\rm DM}h^2$ is shown by dotted black line.
One can see that $\Omega_Ah^2$ is less than the observed value up to $m_A\simeq$ a few TeV,   depending on $\alpha$.
Note that $\Omega_Ah^2$ is highly suppressed at $m_A\simeq m_{H_1}/2$ regardless of $\alpha$, which is due to the resonant enhancement of the annihilation cross section:
\begin{align}
\lefteqn{\sigma(AA\to H_{1,2}\to XX)}\nonumber \\ 
&\propto 
\left|
	\frac{\lambda_{H_1AA}g^{}_{H_1XX}}{s-m_{H_1}^2+im_{H_1}\Gamma_{H_1}} 
	+\frac{\lambda_{H_2AA}g^{}_{H_2XX}}{s-m_{H_2}^2+im_{H_2}\Gamma_{H_2}}	
\right|^2 ~,
\end{align}
where $X$ denotes the gauge bosons or fermions, 
$\Gamma_{H_{1,2}}$ are the total widths of $H_{1,2}$, and $s\simeq 4m_A^2$.
The Higgs couplings are, respectively, given by 
\begin{align}
\lambda_{H_1AA} &= (\delta_2v_0\cos\alpha+d_2v_{S0}\sin\alpha)/2, \\
\lambda_{H_2AA} &= (-\delta_2v_0\sin\alpha+d_2v_{S0}\cos\alpha)/2,\\
g_{H_1VV}^{} &= g_{H_1\bar{f}f}^{} = \cos\alpha, \\
g_{H_2VV}^{} &= g_{H_2\bar{f}f}^{} = -\sin\alpha.
\end{align}
Unlike the real singlet DM scenario, one sees more dips in the curves of Fig.~\ref{fig:DM_nonInert} due to the contribution of $H_2$.

The right plot shows $\tilde{\sigma}_{\rm SI}^{ N}$ as a function of $m_A$. The style and color schemes of the curves are the same as those in the left plot.
The dotted black curve is the LUX exclusion bound.
The allowed regions are $m_A\gtrsim 475$ GeV for $\alpha=-15.0^\circ$, $m_A\gtrsim 1560$ GeV for $\alpha=-20.5^\circ$, $m_A\gtrsim 1995$ GeV for $\alpha=-22.0^\circ$,
and at $m_A\simeq m_{H_1}/2$ for the three cases.

It should be emphasized that even though the large $m_A$ regions are consistent with the DM data,
the SFOEWPT is not realized since the $T=0$ vacuum energy of phase A is lower than that of phase B in this regime,
as mentioned in the last paragraph of the previous section.
For $\alpha=-20.5^\circ$, the occurrence of a SFOEWPT leads to an upper bound of $m_A\lesssim 700$ GeV.

\begin{figure*}
\center
\includegraphics[width=7cm]{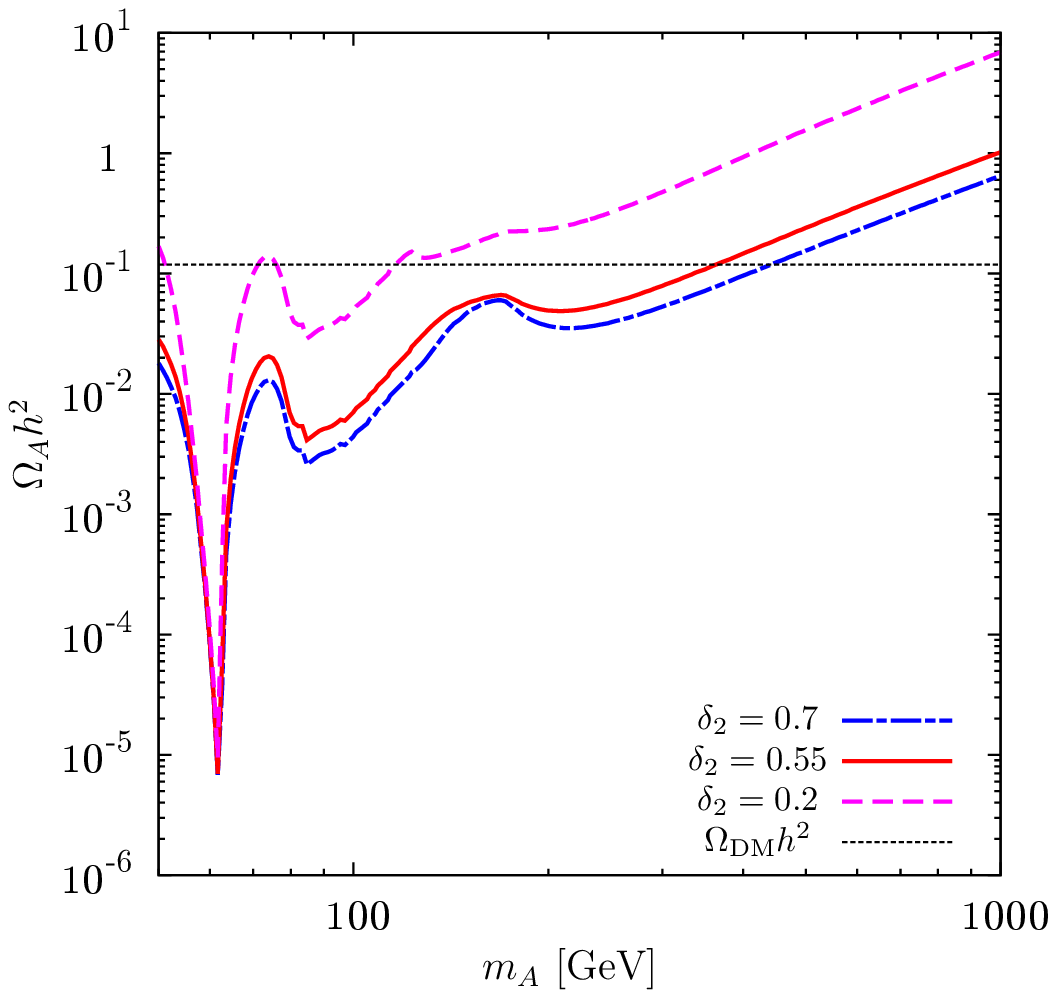}
\hspace{1cm}
\includegraphics[width=7.1cm]{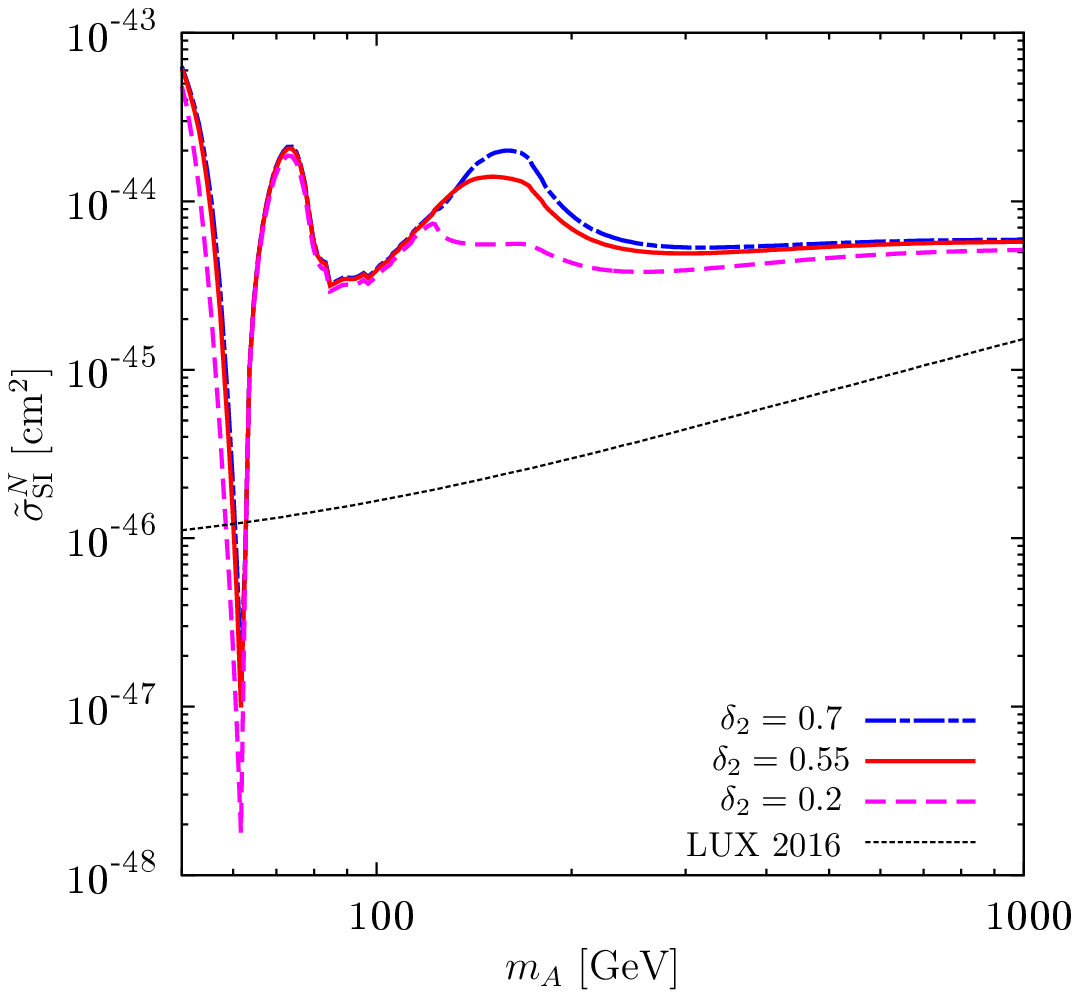}
\caption{Scalar dark matter for case S2: (Left) Relic density of $A$ and $S$ with $\delta_2=0.7$, 0.55 and 0.2, where
$m_A=m_S$ is assumed.
(Right) Scaled spin-independent DM-nucleon scattering cross section.}
\label{fig:DM_Inert}
\end{figure*}
The quantities  $\Omega_Ah^2$ and $\tilde{\sigma}_{\rm SI}^{ N}$ in the case of S2 are shown
 in the left and right plots of Fig.~\ref{fig:DM_Inert}, respectively.
In this case, the scalar particle $H_2(=S)$ is also stable. 
We consider the illustrative situation in which $m_{H_2}=m_A$, and thus both particles contribute to the relic density (for a more general discussion of two-component DM in the cxSM, see Refs.~\cite{Barger:2008jx,Bhattacharya:2016ysw}).
We take $\delta_2=0.7$ (blue, dot-dashed), 0.55 (red, solid) and 0.2 (magenta, dashed).
One can see that a larger $\delta_2$ gives a smaller $\Omega_Ah^2$. 
In contrast to the S1 case, there are regions below 1~TeV for which  
$\Omega_Ah^2=0.1186$ in all three cases.
However, such points are disfavored by the LUX data, as shown in the right plot.
The reason why S2 is more constrained than S1 by the DM direct detection 
is that the scaling factor $\Omega_A/\Omega_{\rm DM}$ defined in Eq.~(\ref{tildeSigmaSI}) 
is larger than that in S1.
We conclude that only the $m_A\simeq 62.5$ GeV case is allowed by the LUX data, 
yet $\Omega_Ah^2$ is well below the observed value. 

Now, we comment on the dependence of $m_A$ on $T_C$, starting with case S2.
To be consistent with the nonzero $\bar{v}_S^{\rm A}(T_C)$ given in Eq.~(\ref{vStilde}), 
$m_A$ must satisfy
\begin{align}
m_A < \sqrt{\frac{\delta_2}{4}v_0^2-2\Sigma_ST_C^2},
\end{align}
where the mass formula (\ref{Hmass_S2}) is used. 
Our numerical study shows that $m_A\lesssim 81.9$ GeV. In the S1 case, Eq.~(\ref{vStilde}) does not apply, while its analog is not particularly enlightening. Consequently, we draw entirely upon a numerical exploration of the $m_A$-dependence and  obtain an upper bound of around 700 GeV for $\alpha = -20.5$.


\section{Phenomenological discussions and comparisons with a previous study}
\label{sec:pheno}
Let us briefly discuss collider phenomenology of our benchmark scenarios.  We do so simply to illustrate the degree to which the parameter space leading to both a viable DM candidate and a SFOEWPT may be probed at the LHC. 
We defer a comprehensive analysis to a future study.  For case S1, $H_2$ can decay into an $AA$ pair, with the partial width given by 
\begin{align}
\Gamma_{H_2\to AA} &= \frac{\lambda_{H_2AA}^2}{32\pi m_{H_2}}\sqrt{1-\frac{4m_A^2}{m_{H_2}^2}} ~.
\end{align}
For $\alpha=-20.5^\circ$, we obtain ${\rm Br}(H_2\to AA)\simeq1$.
In the narrow width approximation, this leads to $\sigma(gg\to H_2){\rm Br}(H_2\to AA)
=s_\alpha^2\sigma(gg\to H_1(230)){\rm Br}(H_2\to AA)\simeq 2.36~(5.42)$~pb, 
where $H_1(230)$ denotes a SM-like Higgs boson with the mass 230 GeV,
and where we have used
$\sigma(gg\to H_1(230))=5.57~(15.1)$~pb at 8 (13)~TeV LHC~\cite{Heinemeyer:2013tqa}. 
With these rates, one could in principle search for an invisibly-decaying heavy Higgs at the LHC. \footnote{Since $A$ is the DM, its typical collider signature is a missing transverse momentum recoil against a visible system. For recent Higgs invisible decay searches at the LHC, see, {\it e.g.}, Ref.~\cite{Khachatryan:2016whc}.}

For case S2, on the other hand, it is difficult to probe at the colliders as it does not have a distinctive signature.  This is because $m_{H_2}=m_A=m_{H_1}/2=62.5$~GeV and the signal strengths of $H_1$ are the same as in the SM. 
However, if the DM mass lies slightly below $m_{H_1}/2$, $H_1$ could have an invisible decay mode, which can in principle be probed via the vector boson fusion processes. 
Detailed studies can be found in Refs.~\cite{Barger:2007im,Barger:2008jx,Feng:2014vea,He:2016mls}.

Before concluding, we make a comparison of this work with Ref.~\cite{Jiang:2015cwa}.
The latter study includes the operators  $H^\dagger H\mathbb{S}$, $\mathbb{S}|\mathbb{S}|^2$
and their Hermitian conjugates that are not considered in the current work. 
On the other hand, the tadpole term $\mathbb{S}$ does not exist in Ref.~\cite{Jiang:2015cwa}.
With those differences in mind, some distinctive conclusions between the two studies emerge: in our analysis,
(1) the one-step EWPT ($\text{O}\to \text{B}$ transition) is absent
and (2) there is no parameter space that can accommodate
both SFOEWPT and the observed DM relic density simultaneously
in the large $m_A$ region.

The former may be attributed to the model setup since the operators 
$H^\dagger H\mathbb{S}$ and $\mathbb{S}|\mathbb{S}|^2$ play an important role
in realizing the one-step EWPT, as argued in Section~\ref{sec:thermal} 
(see also Ref.~\cite{Jiang:2015cwa}).
On the other hand, the latter may be due to the method used in analyzing the EWPT.
At tree level, the approaches are equivalent. Indeed, as discussed in Ref.~\cite{Jiang:2015cwa}, $m_A$ is irrelevant to the vacuum energy difference
between phases A and B at the tree level, which implies that the two-step EWPT is independent
of $m_A$. Beyond this order,  the vacuum energies are no longer independent of $m_A$. Consequently, differences in treating the EWPT beyond tree level will have implications for the viability of a SFOEWPT as a function of $m_A$. Specifically, 
in our gauge-invariant NLO calculation, a large $m_A$ is not compatible with a SFOEWPT
since the $T=0$ vacuum energy of phase A lies below that of phase B as discussed above. This situation represents a 
clear difference from the findings in Ref.~\cite{Jiang:2015cwa}. 

\section{Conclusions}
\label{sec:conclude}
The complex scalar extension of the SM, or cxSM, is a  minimal scenario with the potential to address two issues in cosmology: the generation of a SFOEWPT as needed for EWBG and particle dark matter. With two additional scalar degrees of freedom, it provides an in-principle viable DM candidate as the Goldstone boson of a spontaneously broken global U(1) symmetry (given a mass with soft symmetry-breaking operators), while allowing the remaining scalar to catalyze a SFOEWPT as in the real scalar singlet extensions of the SM.  
The EWPT and DM dynamics are intertwined, governed by a common set of gauge-invariant operators.  

The cxSM also provides a tractable framework for addressing several theoretical issues that pertain to the dynamics of electroweak symmetry-breaking transition. The purpose of the present study is to determine the degree to which a careful treatment of these issues, enumerated in Section \ref{sec:intro}, would affect the viability of the cxSM for both DM and baryogenesis. In general, we find that the use of a RG-improved potential can be implemented in a way that essentially eliminates otherwise problematic scale-dependence while allowing for a gauge-invariant treatment of the EWPT. For the cxSM, this gauge-invariant, RG-improved treatment generally leads to a restricted region of the model parameter space that is consistent with the results of DM direct detection experiments and a two-step scenario for a SFOEWPT. For the basis of operators included in our study, the two-step SFOEWPT is viable only for a finely-tuned region of parameter space where the DM mass is roughly half that of the SM-like Higgs scalar. It is possible that inclusion of cubic U(1)-breaking operators could enable a single-step SFOEWPT with a wider range of DM masses, though an analysis of this possibility should be revisited using a gauge-invariant, RG-improved framework.

Our study also illustrates the need to exercise caution when applying perturbation theory to EWPT dynamics, a principle that goes beyond the cxSM. \footnote{This point has been emphasized in Ref.~\cite{Patel:2011th} and more recently in Ref.~\cite{Curtin:2016urg}, which develops a refined thermal resummation method.} 
We have observed quantitatively noticeable differences between two gauge-invariant approaches: the high-temperature effective theory in which one retains only the finite-$T$ quadratic terms, and the $\hbar$-expansion. Use of the latter appears to imply more effective preservation of the baryon asymmetry than does the former. These differences can become more pronounced near the critical temperature as well as near the endpoints of a first order transition in various regions of parameter space, regions where one might expect higher order contributions to be relatively more important. The assumption of a $T$-independent proportionality between the leading-order sphaleron energy and the gauge-invariant scale associated with EWSB can also break down in the vicinity of $T_C$. We anticipate that these lessons will apply to EWPT dynamics in other models, pointing to the importance of developing more refined perturbative treatments, comparing with non-perturbative computations for representative benchmark parameter choices, and avoiding overly strong conclusions regarding the viability of a given model.

\begin{acknowledgments}
 CC and ES were supported in part by the Ministry of Science and Technology of Taiwan under Grant Nos. 104-2628-M-008-004-MY4 and 105-2811-M-002-126, respectively, and IBS under the project code, IBS-R018-D1 (ES). MJRM was supported in part by U.S. Department of Energy contract DE-SC0011095. MJRM is also grateful for the hospitality of the Department of Physics at the University of Arizona, where a portion of this work was completed. 
\end{acknowledgments}

\appendix

\section{Sphaleron solutions}\label{app:Esph}

To compute $E_{\rm sph}(T)$,  we obtain the sphaleron solutions following a method suggested in Refs.~\cite{Manton:1983nd,Klinkhamer:1984di} and its extension including a singlet scalar field~\cite{Funakubo:2005bu}.
We neglect the effects of U$(1)_Y$ as they are less than a few percent~\cite{Klinkhamer:1990fi,James:1992re}.

A noncontractible loop configuration is given by
\begin{align}
\lefteqn{U(\mu,\theta,\phi)} \nonumber\\
&=
\left(
	\begin{array}{cc}
	e^{i\mu}(\cos\mu-i\sin\mu\cos\theta) & e^{i\phi}\sin\mu\sin\theta \\
	-e^{-i\phi}\sin\mu\sin\theta & e^{-i\mu}(\cos\mu+i\sin\mu\cos\theta)
	\end{array}
\right),
\end{align}
where $\mu$ runs from 0 to $\pi$, parameterizing a least energy path between two adjacent topologically distinct vacua.
The configuration at $\mu=\pi/2$ corresponds to the sphaleron.

The gauge and Higgs fields are expressed in terms of the above noncontractible loop configuration as
\begin{align}
 A_i(\mu,r,\theta,\phi)&=\frac{i}{g_2}f(r)\partial_iU(\mu,\theta,\phi)U^{-1}(\mu,\theta,\phi),\\ 
\Phi(\mu,r,\theta,\phi)&=\frac{\bar{v}(T)}{\sqrt{2}}\bigg[(1-h(r))
	\left(
	\begin{array}{c}
	0 \\
	e^{-i\mu}\cos\mu
	\end{array}
	\right) \nonumber \\
&\hspace{1.5cm}+h(r)U(\mu,\theta,\phi)
	\left(
	\begin{array}{c}
	0 \\
	1
	\end{array}
	\right)\bigg], \\
S(\mu, r, \theta, \phi) &= \frac{\bar{v}_S^{\text{B}}(T)}{\sqrt{2}}k(r),
\end{align}
where $\bar{v}$ and $\bar{v}_S$ are determined using Eq.~(\ref{V_highT_cxSM}).

The energy functional in the $A_0=0$ gauge takes the form
\begin{align}
E_{\rm sph}[f,h,k;T] = \frac{4\pi \Omega(T)}{g_2}\mathcal{E}(T)~,
\end{align}
where
\begin{align}
\mathcal{E}(T)&=\int^\infty_0 d\xi
\bigg[
	4f'^2+\frac{8}{\xi^2}(f-f^2)^2
	+\frac{\xi^2}{2}\frac{v^2}{\Omega^2}h'^2 \nonumber\\
&\hspace{2.5cm}
	+\frac{v^2}{\Omega^2}h^2(1-f)^2
	+\frac{\xi^2}{2}\frac{v_S^2}{\Omega^2}k'^2 
	\nonumber\\
&\hspace{2.5cm}
	+\frac{\xi^2}{g_2^2\Omega^4}V^{\text{high-}T}(h,k;T)
\bigg],
\end{align}
with $\xi=g_2 \Omega r$.
The prime on them denotes a derivative with respect to $\xi$.  
From the energy functional, one can find the equations of motion for the sphaleron configurations:
\begin{align}
\begin{split}
\frac{d^2f}{d\xi^2} &=\frac{2}{\xi^2}(f-f^2)(1-2f)-\frac{v^2 h^2}{4\Omega^2}(1-f),\\
\frac{d}{d\xi}\left[\xi^2\frac{dh}{d\xi}\right]
&= 2h(1-f)^2+\frac{\xi^2}{g_2^2}\frac{1}{\bar{v}^2\Omega^2}\frac{\partial V^{\text{high-}T}}{\partial h}, \\
\frac{d}{d\xi}\left[\xi^2\frac{dk}{d\xi}\right]
&= \frac{\xi^2}{g_2^2}\frac{1}{\bar{v}_S^2\Omega^2}\frac{\partial V^{\text{high-}T}}{\partial k},
\end{split}
\end{align}
with the boundary conditions:
\begin{align}
\begin{split}
\lim_{\xi\to0} f(\xi) &= 0,\quad \lim_{\xi\to0} h(\xi) = 0,\quad \lim_{\xi\to0} k'(\xi) = 0,  \\
\lim_{\xi\to\infty} f(\xi) &= 1,\quad \lim_{\xi\to\infty} h(\xi) = 1,\quad \lim_{ \xi\to \infty} k(\xi) = 1.
\end{split}
\end{align}

\bibliographystyle{h-physrev3.bst}
\bibliography{cxSMrefs}

\end{document}